\documentclass{svjour3}
\usepackage{graphicx}

\begin{document}

\title{Galactic tide and secular orbital evolution}
\author{P. P\'{a}stor \and J. Kla\v{c}ka \and L. K\'{o}mar}
\institute{P. P\'{a}stor \and J. Kla\v{c}ka \and L. K\'{o}mar \at
      Department of Astronomy, Physics of the Earth, and Meteorology, \\
      Faculty of Mathematics, Physics and Informatics, \\
      Comenius University,
      Mlynsk\'{a} dolina, 842~48 Bratislava, Slovak Republic \\
      \email{\{pavol.pastor,klacka,komar\}@fmph.uniba.sk}
}

\date{}

\authorrunning{P. P\'{a}stor et al.}
\titlerunning{Galactic tide and secular orbital evolution}
\maketitle

\begin{abstract}
Equation of motion for the galactic tide is treated
for the case of a comet situated in the Oort cloud of comets.
We take into account that galactic potential and mass density depend
on a distance from the galactic equator and on a distance
from the rotational axis of the Galaxy. Secular evolution of orbital
elements is presented. New terms generated by the Sun's oscillation
about the galactic plane are considered. The inclusion of
the new terms into the equation of motion of the comet leads to
orbital evolution which may be significantly different from the conventional
approach. The usage of the secular time derivatives is limited
to the cases when orbital period of the comet is much less than i)
the period of oscillations of the Sun around the galactic equator,
and, ii) the orbital period of the motion of the Sun around
the galactic center.

\end{abstract}

\section{Introduction}

Global galactic gravitational field influences motion of a comet in the
Oort cloud in the form of the galactic tide.
The motion of the comet with respect to the Sun is important in better
understanding of the Oort cloud.
This paper presents equations for secular evolution of orbital elements
for the comet under the gravity of the Sun and the galactic tide.
We consider equation of motion derived in Kla\v{c}ka (2009a).
Results of our paper reduce to the results
obtained by Kla\v{c}ka and Gajdo\v{s}\'{\i}k (1999) and
Fouchard et al. (2005) when several physical terms are ignored.
The results are compared with detailed numerical solution of the equation
of motion given by K\'{o}mar et al. (2009).

\section{Equation of motion}

We are interested in motion of a comet with respect to the Sun.
The comet is in the Oort cloud and we want to describe the cometary evolution
in terms of secular evolution of comet's orbital elements.

The Sun is moving at a distance $R_{0}$ $=$ 8 kpc from the center
of the Galaxy. Currently, the Sun is situated 30 pc above the galactic
equatorial plane ($Z_{0}$ $=$ 30 pc). Besides rotational motion with
the speed ($A$ $-$ $B$) $\times$ $R_{0}$ (where $A$ and $B$ are Oort
constants) the Sun is moving with the speed 7.3 km/s in the direction
normal to the galactic plane. Positional vector of the comet with respect
to the Sun is $\vec{r}$ $=$ ($\xi$, $\eta$, $\zeta$).
Equation of motion is taken in the form
\begin{eqnarray}\label{1}
\frac{d^{2} \xi}{dt^{2}} &=& - ~ \frac{G M_{\odot}}{r^{3}} ~ \xi ~+~ ( A - B )
\left [ A + B + 2 A \cos \left ( 2 ~ \omega_{0} t \right ) \right ] ~ \xi
\nonumber \\
& & -~ 2 A ( A - B ) \sin \left ( 2 ~ \omega_{0} t \right ) ~\eta
\nonumber \\
& & +~ 2 ( A - B )^{2} ~( \Gamma_{1} - \Gamma_{2} Z_{0}^{2} ) ~
R_{0} ~Z_{0} ~\cos \left ( \omega_{0} t \right ) ~\zeta ~,
\nonumber \\
\frac{d^{2} \eta}{dt^{2}} &=& - ~ \frac{G M_{\odot}}{r^{3}} ~ \eta
~-~ 2 A ( A - B ) \sin \left ( 2 ~ \omega_{0} t \right ) ~ \xi
\nonumber \\
& & +~ ( A - B ) \left [ A + B - 2 A \cos \left ( 2 ~ \omega_{0} t \right )
\right ] ~ \eta
\nonumber \\
& & -~ 2 ( A - B )^{2} ~( \Gamma_{1} - \Gamma_{2} Z_{0}^{2} ) ~
R_{0} ~Z_{0} ~\sin \left ( \omega_{0} t \right ) ~\zeta ~,
\nonumber \\
\frac{d^{2} \zeta}{dt^{2}} &=& - ~ \frac{G M_{\odot}}{r^{3}} ~ \zeta
~-~ \left [ 4 ~\pi ~G ~\varrho ~+~
2 \left ( A^{2} ~-~ B^{2} \right ) \right ] ~\zeta
\nonumber \\
& & -~ 4 ~\pi ~G ~\varrho' ~
    Z_{0} \left [ \cos \left ( \omega_{0} t \right ) ~ \xi ~-~
\sin \left ( \omega_{0} t \right ) ~ \eta \right ] ~,
\nonumber \\
\frac{d^{2} Z_{0}}{dt^{2}} &=& -~ \left [ 4 ~\pi ~G ~\varrho ~+~
2 \left ( A^{2} ~-~ B^{2} \right ) \right ] ~Z_{0} ~,
\nonumber \\
r &=& \sqrt{\xi ^{2} ~+~ \eta ^{2} ~+~ \zeta ^{2}} ~,
\nonumber \\
\omega_{0} &=& A ~-~ B ~,
\end{eqnarray}
where $G$ is the gravitational constant, $M_{\odot}$ is the mass
of the Sun and
\begin{eqnarray}\label{2}
A &=& 14.2 ~\mbox{km} ~\mbox{s}^{-1} ~\mbox{kpc}^{-1} ~,
\nonumber \\
B &=& - 12.4 ~\mbox{km} ~\mbox{s}^{-1} ~\mbox{kpc}^{-1} ~,
\nonumber \\
\Gamma_{1} &=& 0.124 ~\mbox{kpc}^{-2} ~,
\nonumber \\
\Gamma_{2} &=& 1.586 ~\mbox{kpc}^{-4} ~,
\nonumber \\
\varrho &=& 0.130 ~\mbox{M}_{\odot} ~\mbox{pc}^{-3} ~,
\nonumber \\
\varrho' &=& - 0.037 ~\mbox{M}_{\odot} ~\mbox{pc}^{-3} ~\mbox{kpc}^{-1} ~,
\end{eqnarray}
see Eqs. (26)-(27) in Kla\v{c}ka (2009a). If one wants to use other values
of the Oort constants $A$ and $B$, he can use Eqs. (22) in Kla\v{c}ka (2009a):
\begin{eqnarray}\label{3}
\varrho &=& \varrho_{disk} ~+~ \varrho_{halo} ~,
\nonumber \\
\varrho_{disk} &=& 0.126 ~\mbox{M}_{\odot} ~\mbox{pc}^{-3} ~,
\nonumber \\
\varrho_{halo} &=& (4 \pi G)^{-1}[ X(Galaxy) + X(disk) + X(bulge) ] ~,
\nonumber \\
X(Galaxy) &\equiv& - (A - B) \times (A + 3 B)
\nonumber \\
X(disk) &=& -~396.90 ~\mbox{km}^{2} ~\mbox{s}^{-2} ~\mbox{kpc}^{-2} ~,
\nonumber \\
X(bulge) &=& -~0.65 ~\mbox{km}^{2} ~\mbox{s}^{-2} ~\mbox{kpc}^{-2} ~,
\end{eqnarray}
The value $X(Galaxy)$ $=$ 611.800 km$^{2}$ s$^{-2}$ kpc$^{-2}$ holds for
$A$ $=$ 14.2~ km ~s$^{-1}$ ~kpc$^{-1}$ and
$B$ $=$ $-$ 12.4 km ~s$^{-1}$ ~kpc$^{-1}$.
Eqs. (22) of Kla\v{c}ka (2009a) can be used.

\begin{figure}[t]
\begin{center}
\includegraphics[height=0.23\textheight]{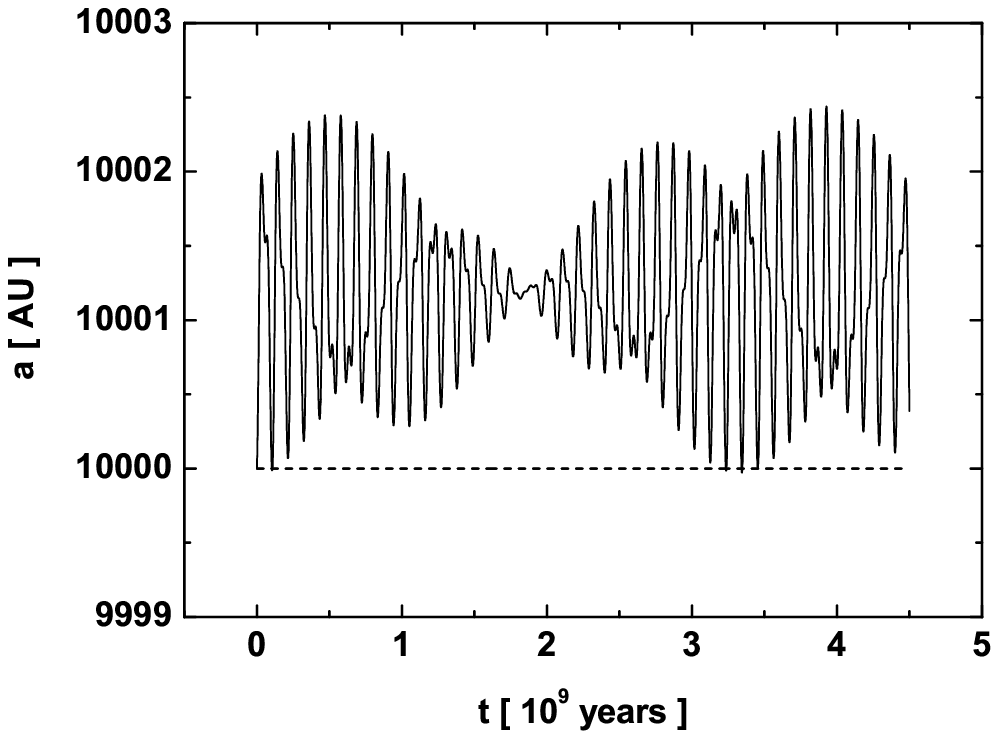}
\includegraphics[height=0.23\textheight]{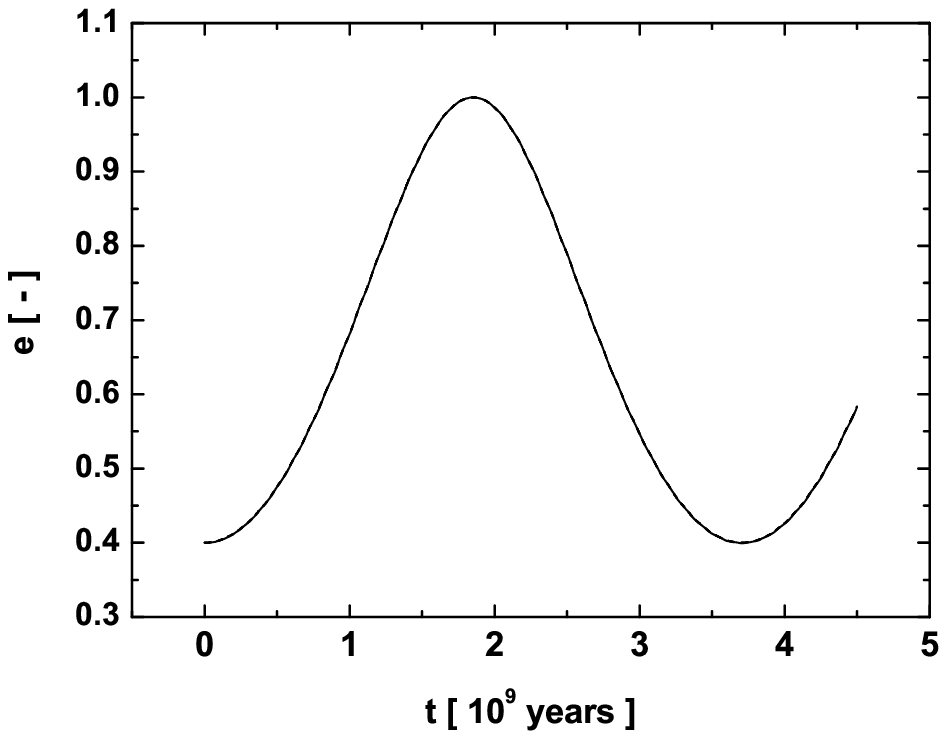}
\includegraphics[height=0.23\textheight]{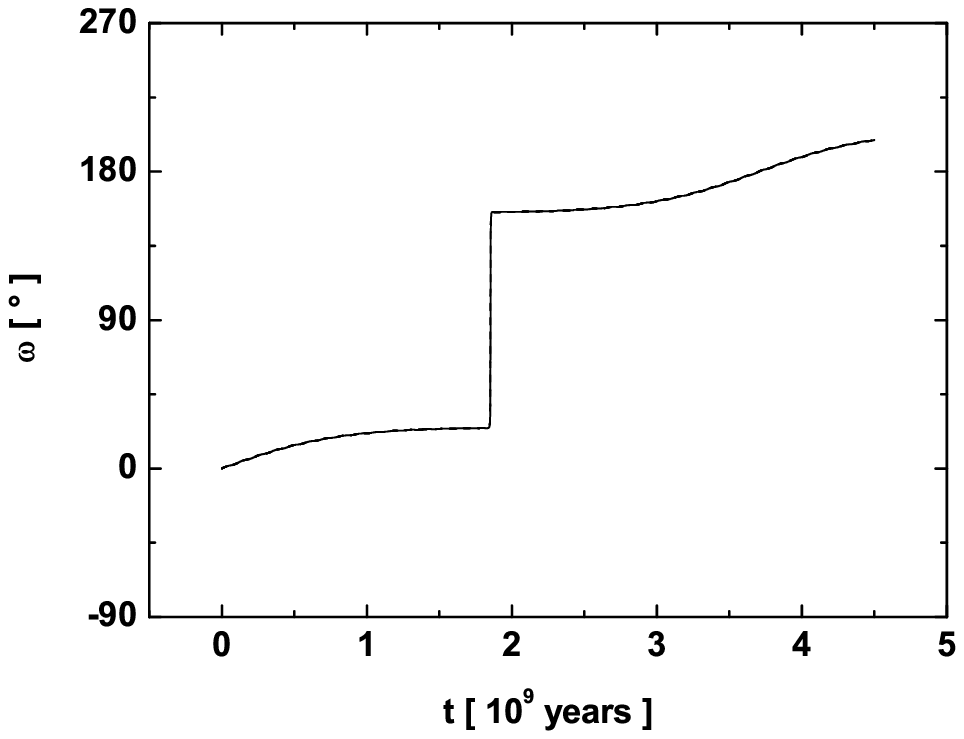}
\includegraphics[height=0.23\textheight]{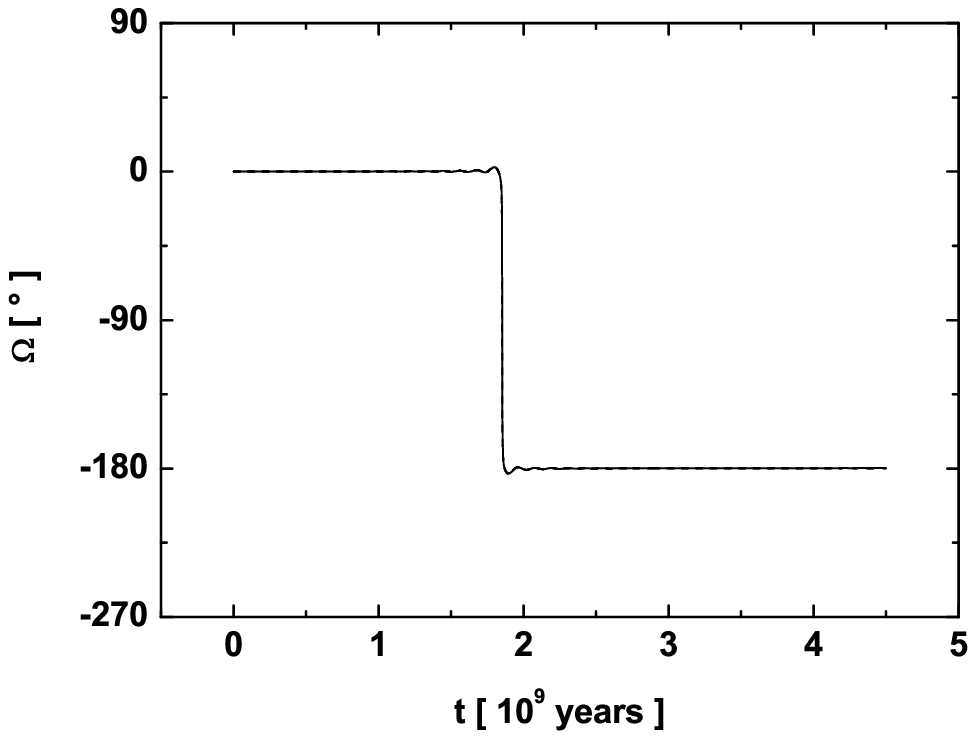}
\includegraphics[height=0.23\textheight]{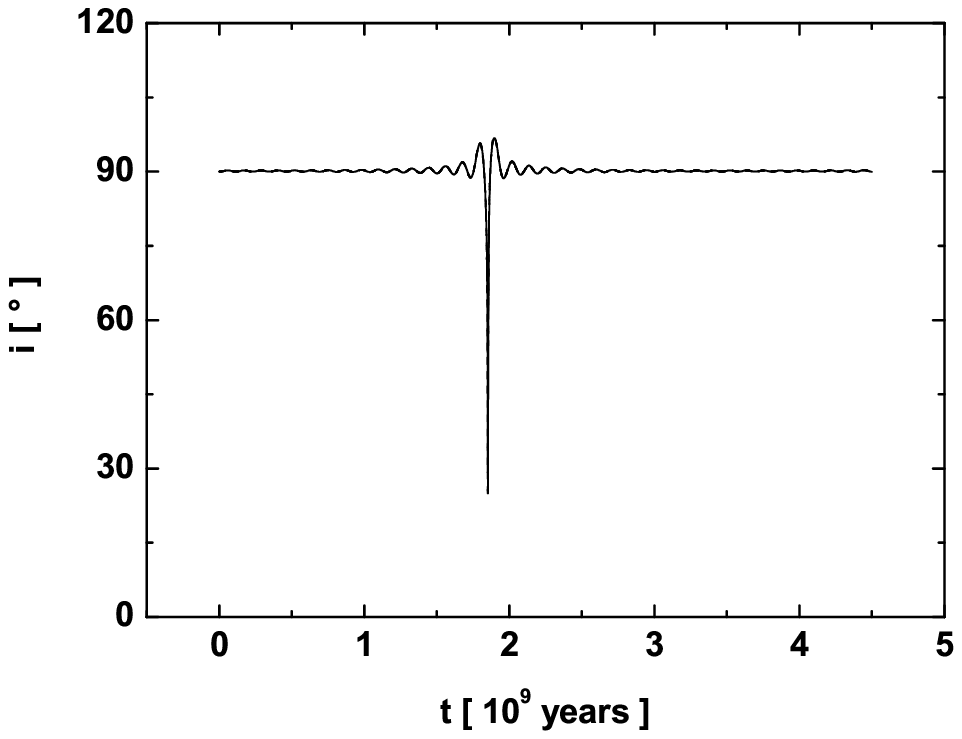}
\end{center}
\caption{Two orbital evolutions of a comet with initial semi-major axis
10000 AU in the Oort cloud under the influence of galactic tide.
Evolutions are obtained from numerical solution of system of differential
equations given by Eqs. (13)-(17) with the new terms (solid line)
and without new terms (dashed line).}
\label{F1}
\end{figure}

\begin{figure}[t]
\begin{center}
\includegraphics[height=0.3\textheight]{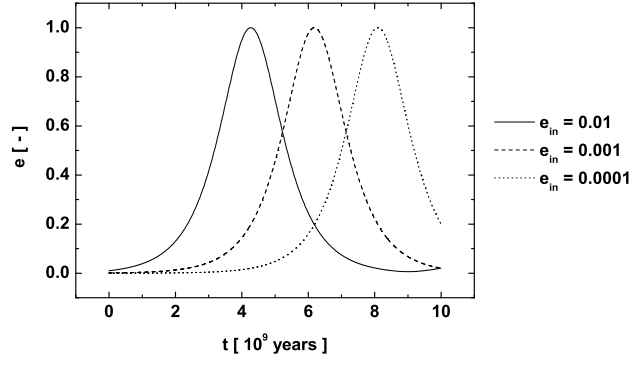}
\end{center}
\caption{Time evolution of eccentricity when initial eccentricity is close
to zero.}
\label{F2}
\end{figure}

\section{Secular changes of orbital elements}

Perturbation equations of celestial mechanics yield for osculating orbital
elements ($a$ -- semi-major axis; $e$ -- eccentricity; $i$ -- inclination
(of the orbital plane to the reference plane -- galactic equatorial plane);
$\Omega$ -- longitude of the ascending node; $\omega$ -- argument
of perihelion; $\Theta$ is the position angle of the particle on the orbit,
when measured from the ascending node in the direction of the particle's
motion, $\Theta = \omega + f$):
\begin{eqnarray}\label{4}
\frac{d a}{d t} &=& \frac{2~a}{1~-~e^{2}} ~
      \sqrt{\frac{p}{\mu}} ~
      \left \{
      F_{R} ~e~ \sin f +
      F_{T} \left ( 1~+~e~ \cos f \right ) \right \} ~,
\nonumber \\
\frac{d e}{d t} &=&
      \sqrt{\frac{p}{\mu}} ~ \left \{F_{R} ~ \sin f + F_{T} \left [ \cos f ~+~
      \frac{e +	\cos f}{1 + e \cos f} \right ] \right \} ~,
\nonumber \\
\frac{d i}{d t} &=& \frac{r}{\sqrt{\mu ~p}} ~F_{N} ~ \cos \Theta ~,
\nonumber \\
\frac{d \Omega}{d t} &=&
      \frac{r}{\sqrt{\mu ~p}} ~
      F_{N} ~ \frac{\sin \Theta}{\sin i} ~,
\nonumber \\
\frac{d \omega}{d t} &=& -~ \frac{1}{e} ~ \sqrt{\frac{p}{\mu}} ~
      \left \{ F_{R} \cos f - F_{T} ~
      \frac{2 + e \cos f}{1 + e \cos f} ~
      \sin f \right \}
\nonumber \\
& & -~ \frac{r}{\sqrt{\mu ~p}} ~
      F_{N} ~ \frac{\sin \Theta}{\sin i} ~\cos i ~,
\nonumber \\
\frac{d \Theta}{d t} &=&
      \frac{\sqrt{\mu ~p}}{r^{2}} ~-~
      \frac{r}{\sqrt{\mu~ p}} ~
      F_{N} ~ \frac{\sin \Theta}{\sin i} ~\cos i ~,
\nonumber \\
r &=& p / ( 1 + e \cos f ) ~,
\nonumber \\
p &=& a~ ( 1 - e^{2} ) ~,
\nonumber \\
\mu &\equiv& G M_{\odot}~,
\end{eqnarray}
where $F_{R}$, $F_{T}$ and $F_{N}$ are radial, transversal and
normal components of the disturbing acceleration.
We use $-~\mu ~\vec{e}_{R}~/~r^{2}$ as a central
acceleration determining osculating orbital elements if we want to take a time
average ($T$ is time interval between passages through two following
pericenters) in an analytical way
\begin{eqnarray}\label{5}
\langle g \rangle &\equiv& \frac{1}{T}	\int_{0}^{T} g(t) dt =
\frac{\sqrt{\mu}}{a^{3/2}} ~
\frac{1}{2 \pi} \int_{0}^{2 \pi} g(f)
\left ( \frac{df}{dt} \right )^{-1} df
\nonumber \\
&=& \frac{\sqrt{\mu}}{a^{3/2}} ~ \frac{1}{2 \pi} \int_{0}^{2 \pi}
g(f) ~\frac{r^{2}}{\sqrt{\mu~p}}~ df
\nonumber \\
&=& \frac{1}{a^{2}~ \sqrt{1 - e^{2}}} ~\frac{1}{2 \pi} ~
  \int_{0}^{2 \pi} ~ g(f) ~r^{2} ~ df ~,
\end{eqnarray}
assuming non-pseudo-circular orbits and the fact that orbital elements exhibit
only small changes during the time interval $T$; the second and the third
Kepler's laws were used: $r^{2} ~df/dt = \sqrt{\mu p}$ -- conservation of
angular momentum, $a^{3}/T^{2} = \mu / (4 \pi^{2})$.

Rewriting Eq. (1) to the form
\begin{eqnarray}\label{6}
\frac{d^{2} \xi}{dt^{2}} &=& - ~ \frac{G M_{\odot}}{r^{3}} ~ \xi
~+~ F_{x} ~,
\nonumber \\
\frac{d^{2} \eta}{dt^{2}} &=& - ~ \frac{G M_{\odot}}{r^{3}} ~ \eta
~+~ F_{y} ~,
\nonumber \\
\frac{d^{2} \zeta}{dt^{2}} &=& - ~ \frac{G M_{\odot}}{r^{3}} ~ \zeta
~+~ F_{z} ~,
\nonumber \\
r &=& \sqrt{\xi ^{2} ~+~ \eta ^{2} ~+~ \zeta ^{2}} ~,
\nonumber \\
F_{x} &=& ( A - B ) \left [ A + B + 2 A \cos \left ( 2 ~ \omega_{0} t \right )
 \right ] ~ \xi
~-~ 2 A ( A - B ) \sin \left ( 2 ~ \omega_{0} t \right ) ~\eta
\nonumber \\
& & +~ 2 ( A - B )^{2} ~( \Gamma_{1} - \Gamma_{2} Z_{0}^{2} ) ~
R_{0} ~Z_{0} ~\cos \left ( \omega_{0} t \right ) ~\zeta ~,
\nonumber \\
F_{y} &=& -~ 2 A ( A - B ) \sin \left ( 2 ~ \omega_{0} t \right ) ~ \xi
~+~ ( A - B ) \left [ A + B - 2 A \cos \left ( 2 ~ \omega_{0} t \right )
\right ] ~ \eta
\nonumber \\
& & -~ 2 ( A - B )^{2} ~( \Gamma_{1} - \Gamma_{2} Z_{0}^{2} ) ~
R_{0} ~Z_{0} ~\sin \left ( \omega_{0} t \right ) ~\zeta ~,
\nonumber \\
F_{z} &=& -~ \omega_{z} ^{2} ~\zeta ~-~ 4 ~\pi ~G ~\varrho' ~
    Z_{0} \left [ \cos \left ( \omega_{0} t \right ) ~ \xi ~-~
    \sin \left ( \omega_{0} t \right ) ~ \eta \right ] ~,
\nonumber \\
Z_{0} &=& K~ \sin ( \omega_{z} ~t + \varphi_{0} ) ~,
\nonumber \\
Z_{0} ( t = t_{0} ) &=& 30 ~\mbox{pc} = 6.188 \times 10^{6} ~\mbox{AU} ~,
\nonumber \\
\dot{Z_{0}} ( t = t_{0} ) &=& 7.3 ~\mbox{km} ~\mbox{s}^{-1} =
1.540 ~\mbox{AU} ~\mbox{yr}^{-1} ~,
\nonumber \\
\omega_{z} &=& \sqrt{4 ~\pi ~G ~\varrho ~+~
2 \left ( A^{2} ~-~ B^{2} \right )} ~,
\nonumber \\
\omega_{0} &=& A ~-~ B ~,
\end{eqnarray}
where $t_{0}$ denotes the current time moment and $\varphi_{0}$ the initial
phase, we can find the required components $F_{R}$, $F_{T}$ and $F_{N}$
of the disturbing acceleration:
\begin{eqnarray}\label{7}
F_{R} &=& \vec{F} \cdot \vec{e}_{R} \equiv
F_{x} ~e_{R~x} ~+~ F_{y} ~e_{R~y} ~+~F_{z} ~e_{R~z} ~,
\nonumber \\
F_{T} &=& \vec{F} \cdot \vec{e}_{T} \equiv
F_{x} ~e_{T~x} ~+~ F_{y} ~e_{T~y} ~+~F_{z} ~e_{T~z} ~,
\nonumber \\
F_{N} &=& \vec{F} \cdot \vec{e}_{N} \equiv
F_{x} ~e_{N~x} ~+~ F_{y} ~e_{N~y} ~+~F_{z} ~e_{N~z} ~,
\nonumber \\
\vec{e}_{R} &=& ( \cos \Omega ~ \cos \Theta ~-~\sin \Omega ~
\sin \Theta ~\cos i,
\nonumber \\
& & \sin \Omega ~ \cos \Theta ~+~\cos \Omega ~ \sin \Theta ~\cos i,
\sin \Theta ~\sin i) ~,
\nonumber \\
\vec{e}_{T} &=& ( -~\cos \Omega ~ \sin \Theta ~-~\sin \Omega ~
\cos \Theta ~\cos i,
\nonumber \\
& & -~ \sin \Omega ~ \sin \Theta ~+~\cos \Omega ~ \cos \Theta ~\cos i,
\cos \Theta ~\sin i) ~,
\nonumber \\
\vec{e}_{N} &=& ( \sin \Omega ~ \sin i, - \cos \Omega ~\sin i, \cos i) ~.
\end{eqnarray}

Inserting Eqs. (6) into Eqs. (7):
\begin{eqnarray}\label{8}
F_{R} &=& \bigl \{( A - B )
\left [ A + B + 2 A \cos \left ( 2 ~ \omega_{0} t \right ) \right ] ~r~
( \cos \Omega ~ \cos \Theta ~-~\sin \Omega ~ \sin \Theta ~\cos i )
\nonumber \\
& & -~ 2 A ( A - B ) \sin \left ( 2 ~ \omega_{0} t \right ) ~r~
( \sin \Omega ~ \cos \Theta ~+~ \cos \Omega ~ \sin \Theta ~\cos i )
\nonumber \\
& &
+~ 2 ( A - B )^{2} ~( \Gamma_{1} - \Gamma_{2} Z_{0}^{2} ) ~
R_{0} ~Z_{0} ~\cos \left ( \omega_{0} t \right ) ~r ~\sin \Theta ~
\sin i \bigr \} ~
\times
\nonumber \\
& & \times ~ ( \cos \Omega ~ \cos \Theta ~-~\sin \Omega ~ \sin \Theta ~\cos i )
\nonumber \\
& & +~ \bigl \{ -~ 2 A ( A - B ) \sin \left ( 2 ~ \omega_{0} t \right ) ~r~
( \cos \Omega ~ \cos \Theta ~-~\sin \Omega ~ \sin \Theta ~\cos i )
\nonumber \\
& & +~ ( A - B ) \left [ A + B - 2 A \cos
\left ( 2 ~ \omega_{0} t \right ) \right ] ~r~
( \sin \Omega ~ \cos \Theta ~+~ \cos \Omega ~ \sin \Theta ~\cos i )
\nonumber \\
& & -~ 2 ( A - B )^{2} ~( \Gamma_{1} - \Gamma_{2} Z_{0}^{2} ) ~
R_{0} ~Z_{0} ~\sin \left ( \omega_{0} t \right ) ~r ~ \sin \Theta ~
\sin i \bigr \} ~
\times
\nonumber \\
& & \times ~ ( \sin \Omega ~ \cos \Theta ~+~\cos \Omega ~ \sin \Theta ~\cos i )
\nonumber \\
& & +~ \bigl \{ -~ \omega_{z} ^{2} ~r~\sin \Theta ~\sin i
\nonumber \\
& & -~ 4 ~\pi ~G ~\varrho' ~
    Z_{0} ~r~ \bigl [ \cos \left ( \omega_{0} t \right ) ~
( \cos \Omega ~ \cos \Theta ~-~\sin \Omega ~ \sin \Theta ~\cos i )
\nonumber \\
& & -~ \sin \left ( \omega_{0} t \right ) ~
( \sin \Omega ~ \cos \Theta ~+~ \cos \Omega ~ \sin \Theta ~\cos i ) \bigr ]
\bigr \}
\times ~ \sin \Theta ~\sin i ~,
\end{eqnarray}

\begin{eqnarray}\label{9}
F_{T} &=& \bigl \{( A - B )
\left [ A + B + 2 A \cos \left ( 2 ~ \omega_{0} t \right ) \right ] ~r~
( \cos \Omega ~ \cos \Theta ~-~\sin \Omega ~ \sin \Theta ~\cos i )
\nonumber \\
& & -~ 2 A ( A - B ) \sin \left ( 2 ~ \omega_{0} t \right ) ~r~
( \sin \Omega ~ \cos \Theta ~+~ \cos \Omega ~ \sin \Theta ~\cos i )
\nonumber \\
& & +~ 2 ( A - B )^{2} ~( \Gamma_{1} - \Gamma_{2} Z_{0}^{2} ) ~
R_{0} ~Z_{0} ~\cos \left ( \omega_{0} t \right ) ~r~ \sin \Theta ~
\sin i \bigr \} ~
\times
\nonumber \\
& & \times ~ ( -~\cos \Omega ~ \sin \Theta ~-~\sin \Omega ~
\cos \Theta ~\cos i )
\nonumber \\
& & +~ \bigl \{ -~ 2 A ( A - B ) \sin \left ( 2 ~ \omega_{0} t \right ) ~r~
( \cos \Omega ~ \cos \Theta ~-~\sin \Omega ~ \sin \Theta ~\cos i )
\nonumber \\
& & +~ ( A - B ) \left [ A + B - 2 A \cos
\left ( 2 ~ \omega_{0} t \right ) \right ] ~ r~
( \sin \Omega ~ \cos \Theta ~+~ \cos \Omega ~ \sin \Theta ~\cos i )
\nonumber \\
& & -~ 2 ( A - B )^{2} ~( \Gamma_{1} - \Gamma_{2} Z_{0}^{2} ) ~
R_{0} ~Z_{0} ~\sin \left ( \omega_{0} t \right ) ~r~ \sin \Theta ~
\sin i \bigr \} ~
\times
\nonumber \\
& & \times ~ ( -~ \sin \Omega ~ \sin \Theta ~+~\cos \Omega ~
\cos \Theta ~\cos i )
\nonumber \\
& & +~ \bigl \{ -~ \omega_{z} ^{2} ~r~\sin \Theta ~\sin i
\nonumber \\
& & -~ 4 ~\pi ~G ~\varrho' ~
    Z_{0} ~r~ \bigl [ \cos \left ( \omega_{0} t \right ) ~
( \cos \Omega ~ \cos \Theta ~-~\sin \Omega ~ \sin \Theta ~\cos i )
\nonumber \\
& & -~ \sin \left ( \omega_{0} t \right ) ~
( \sin \Omega ~ \cos \Theta ~+~ \cos \Omega ~ \sin \Theta ~\cos i ) \bigr ]
\bigr \}
\times ~ \cos \Theta ~\sin i ~,
\end{eqnarray}

\begin{eqnarray}\label{10}
F_{N} &=& \bigl \{( A - B )
\left [ A + B + 2 A \cos \left ( 2 ~ \omega_{0} t \right ) \right ] ~r~
( \cos \Omega ~ \cos \Theta ~-~\sin \Omega ~ \sin \Theta ~\cos i )
\nonumber \\
& & -~ 2 A ( A - B ) \sin \left ( 2 ~ \omega_{0} t \right ) ~r~
( \sin \Omega ~ \cos \Theta ~+~ \cos \Omega ~ \sin \Theta ~\cos i )
\nonumber \\
& & +~ 2 ( A - B )^{2} ~( \Gamma_{1} - \Gamma_{2} Z_{0}^{2} ) ~
R_{0} ~Z_{0} ~\cos \left ( \omega_{0} t \right ) ~r~ \sin \Theta ~
\sin i \bigr \} ~\times ~ ( \sin \Omega ~ \sin i )
\nonumber \\
& & +~ \bigl \{ -~ 2 A ( A - B ) \sin \left ( 2 ~ \omega_{0} t \right ) ~r~
( \cos \Omega ~ \cos \Theta ~-~\sin \Omega ~ \sin \Theta ~\cos i )
\nonumber \\
& & +~ ( A - B ) \left [ A + B - 2 A \cos
\left ( 2 ~ \omega_{0} t \right ) \right ] ~r~
( \sin \Omega ~ \cos \Theta ~+~ \cos \Omega ~ \sin \Theta ~\cos i )
\nonumber \\
& & -~ 2 ( A - B )^{2} ~( \Gamma_{1} - \Gamma_{2} Z_{0}^{2} ) ~
R_{0} ~Z_{0} ~\sin \left ( \omega_{0} t \right ) ~r~ \sin \Theta ~ \sin i
\bigr \} ~ \times ~ ( - \cos \Omega ~\sin i )
\nonumber \\
& & +~ \bigl \{ -~ \omega_{z} ^{2} ~r~ \sin \Theta ~ \sin i
\nonumber \\
& & -~ 4 ~\pi ~G ~\varrho' ~
    Z_{0} ~r~ \bigl [ \cos \left ( \omega_{0} t \right ) ~
( \cos \Omega ~ \cos \Theta ~-~ \sin \Omega ~ \sin \Theta ~\cos i )
\nonumber \\
& & -~ \sin \left ( \omega_{0} t \right ) ~
( \sin \Omega ~ \cos \Theta ~+~ \cos \Omega ~ \sin \Theta ~\cos i ) \bigr ]
\bigr \} \times ~ \cos i ~.
\end{eqnarray}

\begin{figure}[t]
\begin{center}
\includegraphics[height=0.23\textheight]{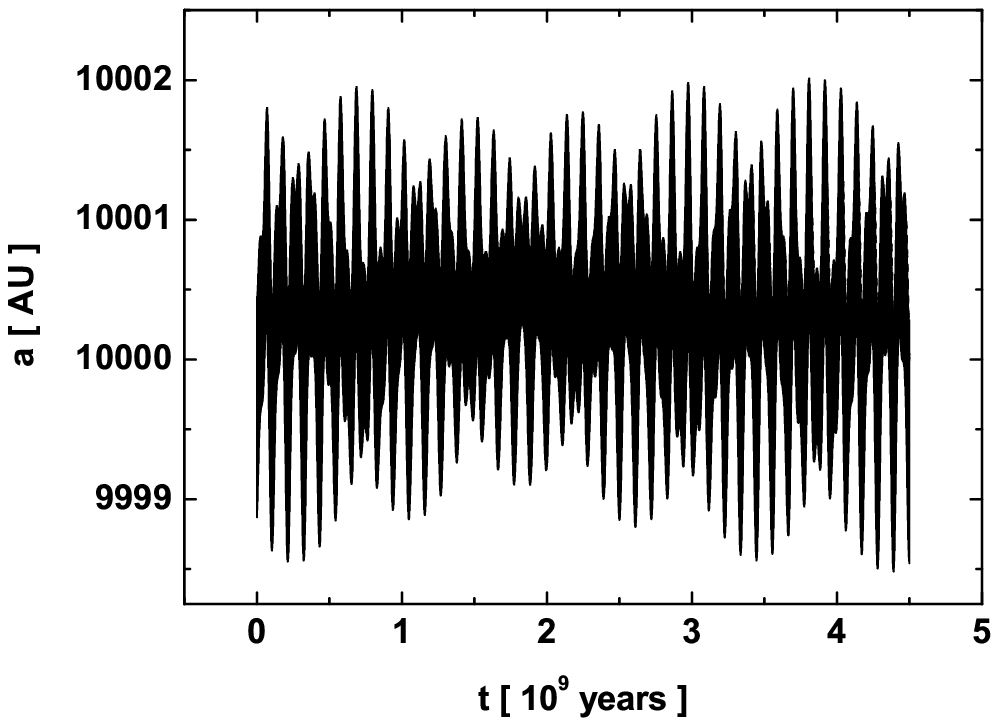}
\includegraphics[height=0.23\textheight]{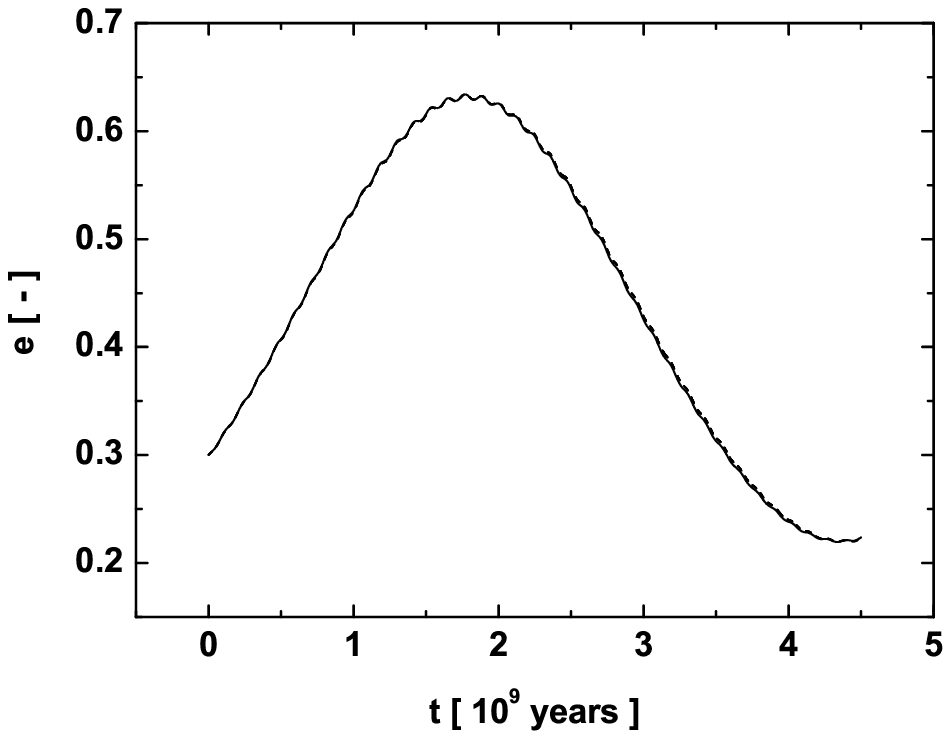}
\includegraphics[height=0.23\textheight]{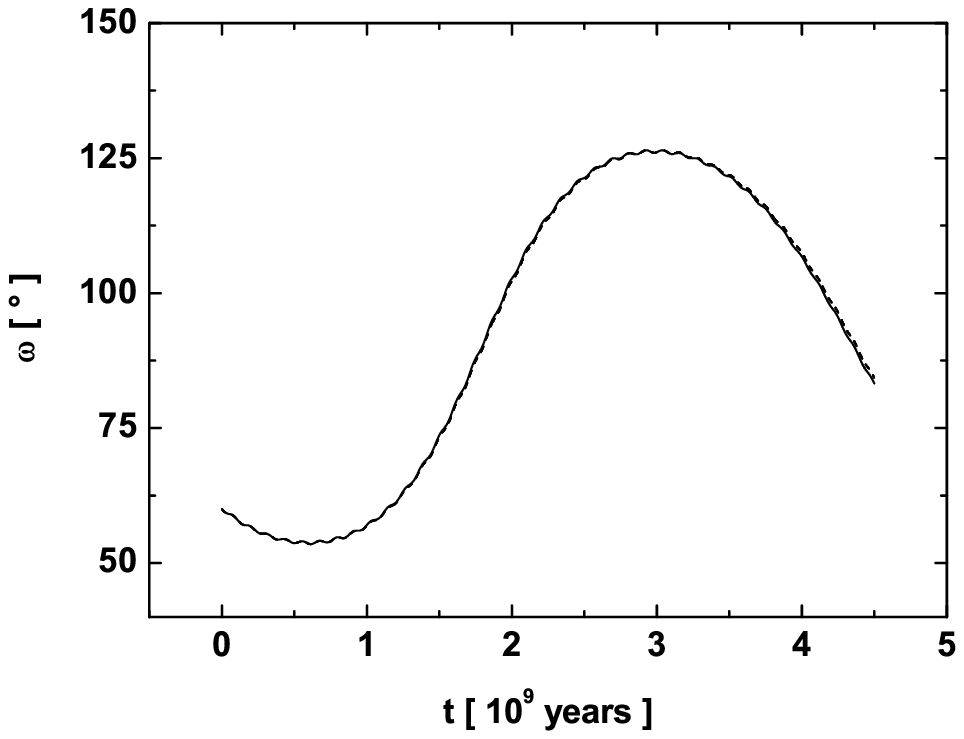}
\includegraphics[height=0.23\textheight]{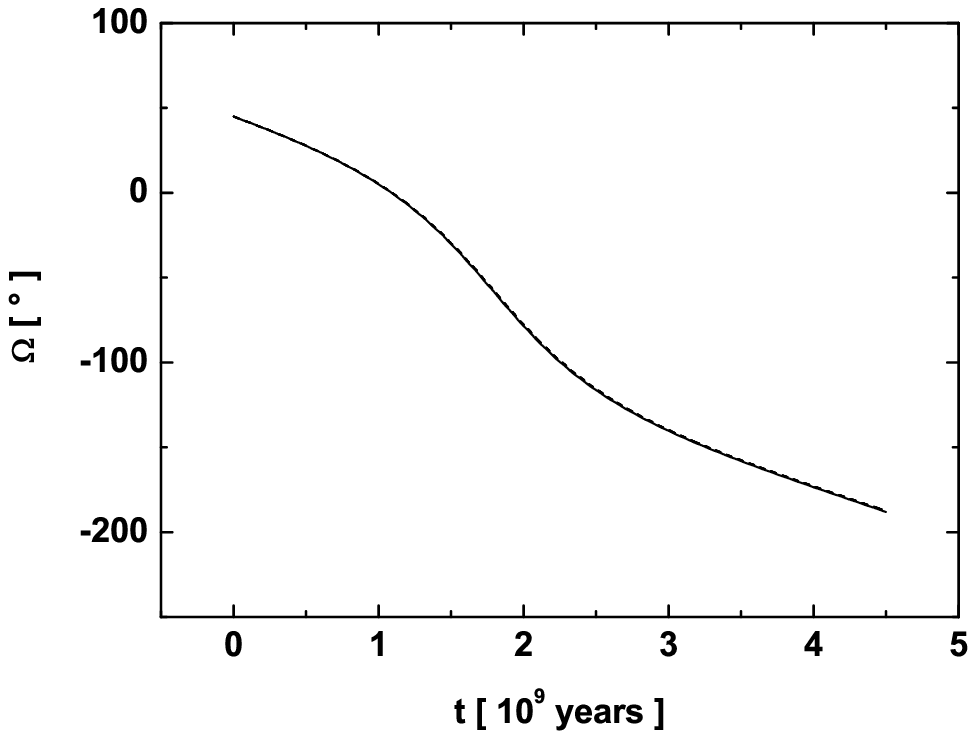}
\includegraphics[height=0.23\textheight]{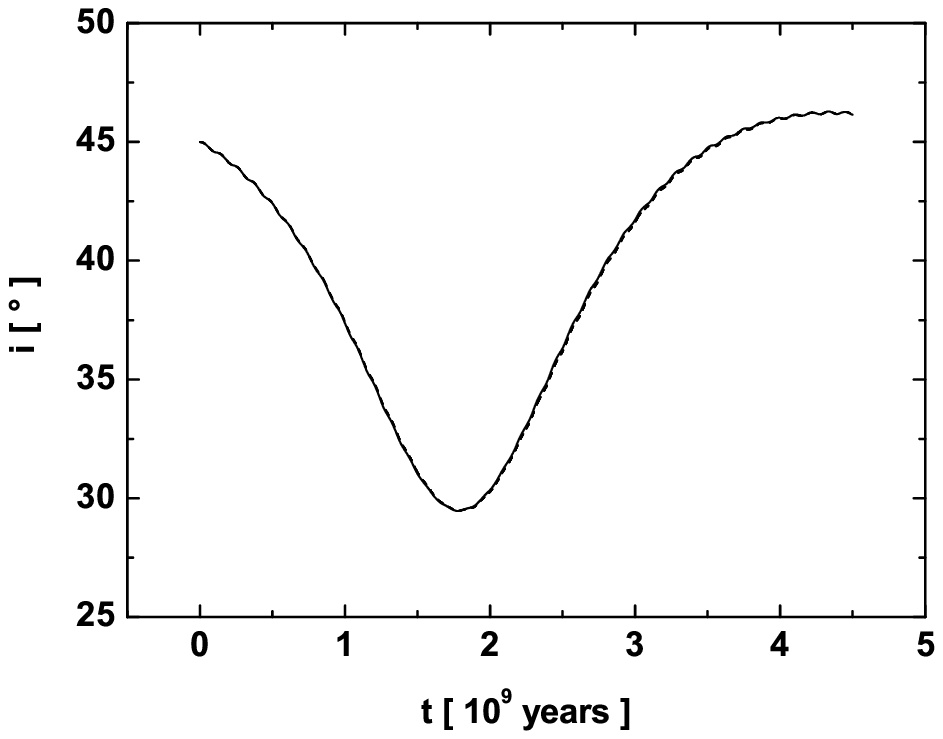}
\end{center}
\caption{Comparison of two orbital evolutions of a comet
with initial semi-major axis 10000 AU obtained by numerical
solution of Eqs. (1) (solid line) and Eqs. (13)-(17)
(dashed line). In both numerical solutions are
the new terms included.}
\label{F3}
\end{figure}

\subsection{A method of averaging}

Inserting Eqs. (8)-(10) into Eqs. (4) and making time averaging
represented by Eq. (5), we need also the relation between the time $t$ and
the true anomaly $f$. The relation $dt$ $=$ ($r^{2} / \sqrt{\mu ~p}$) $df$
yields
\begin{eqnarray}\label{11}
t &=& \tau ~+~ \frac{a^{3/2}}{\sqrt{\mu}} ~\left \{
      2 ~\arctan \left ( \sqrt{\frac{1 - e}{1 + e}} ~
      \tan \frac{f}{2} \right ) ~-~
      e ~ \sqrt{1 - e^{2}} ~\frac{\sin f}{1 + e \cos f} \right \} ~,
\end{eqnarray}
if we take $f (t = \tau) =$ 0.

Another possibility is to use the Kepler's equation
$t = \tau + (a^{3/2} / \sqrt{\mu}) ( E ~-~ e~ \sin E )$
[see also Eq. (11) together with $\tan (E/2)$ $=$
$\sqrt{(1 - e)/(1 + e)}$ $\tan (f/2)$]. Then, instead of Eq. (5), we obtain
\begin{eqnarray}\label{12}
\langle g \rangle &=& \frac{1}{2 \pi} ~
  \int_{0}^{2 \pi} ~ g(E) ~(1 ~-~ e~ \cos E ) ~dE ~,
\nonumber \\
\sin f &=& \frac{a~ \sqrt{1~-~e^{2}}}{r} ~\sin E ~,
\nonumber \\
\cos f &=& \frac{a}{r} ~\left ( \cos E ~-~e \right ) ~,
\nonumber \\
r &=& a~ ( 1 ~-~e ~ \cos E ) ~,
\nonumber \\
t &=& \tau + (a^{3/2} / \sqrt{\mu}) ( E ~-~ e~ \sin E ) ~.
\end{eqnarray}

\begin{figure}[t]
\begin{center}
\includegraphics[height=0.23\textheight]{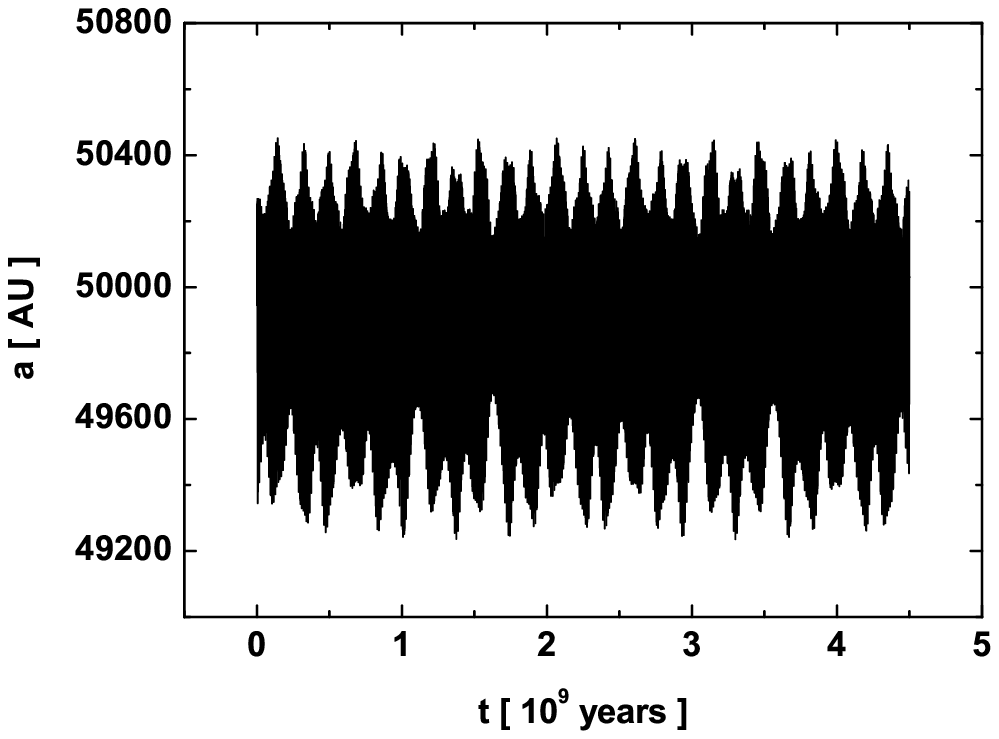}
\includegraphics[height=0.23\textheight]{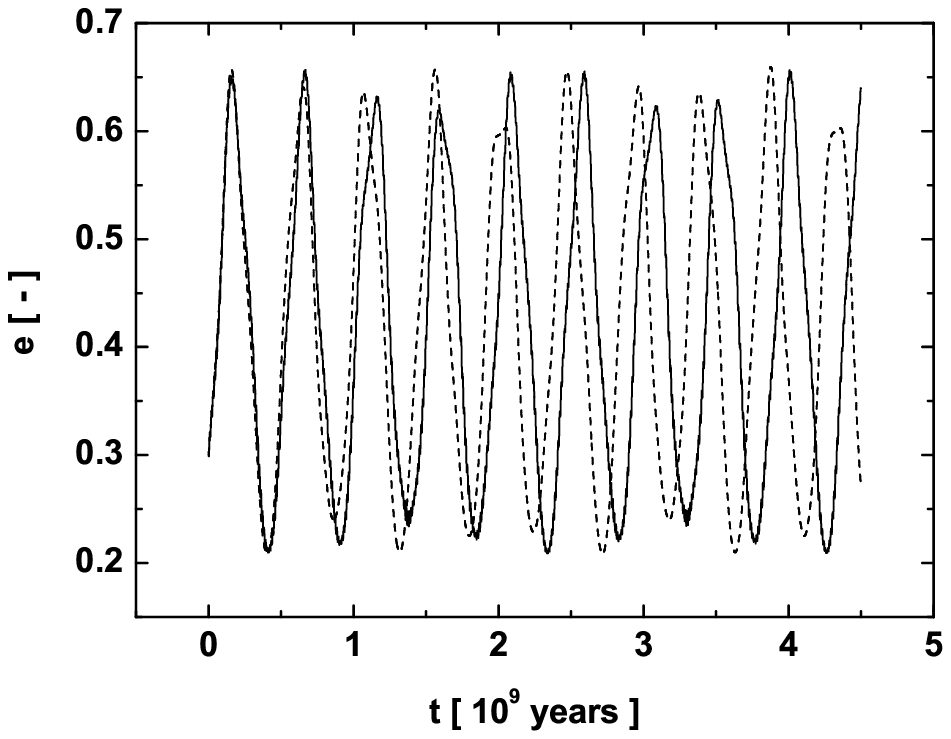}
\includegraphics[height=0.23\textheight]{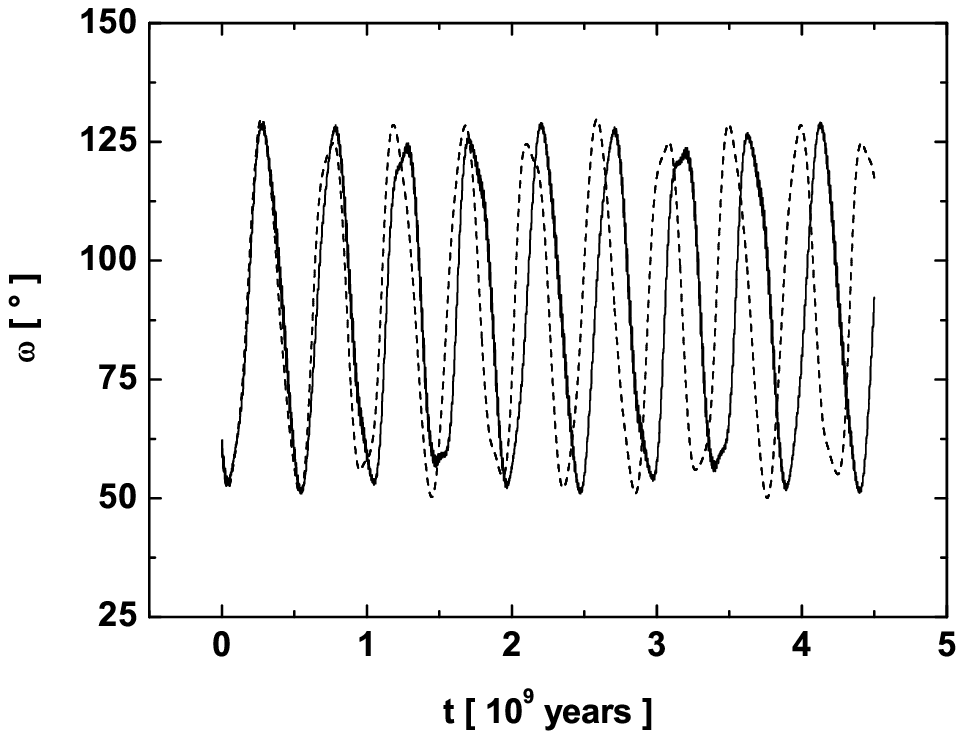}
\includegraphics[height=0.23\textheight]{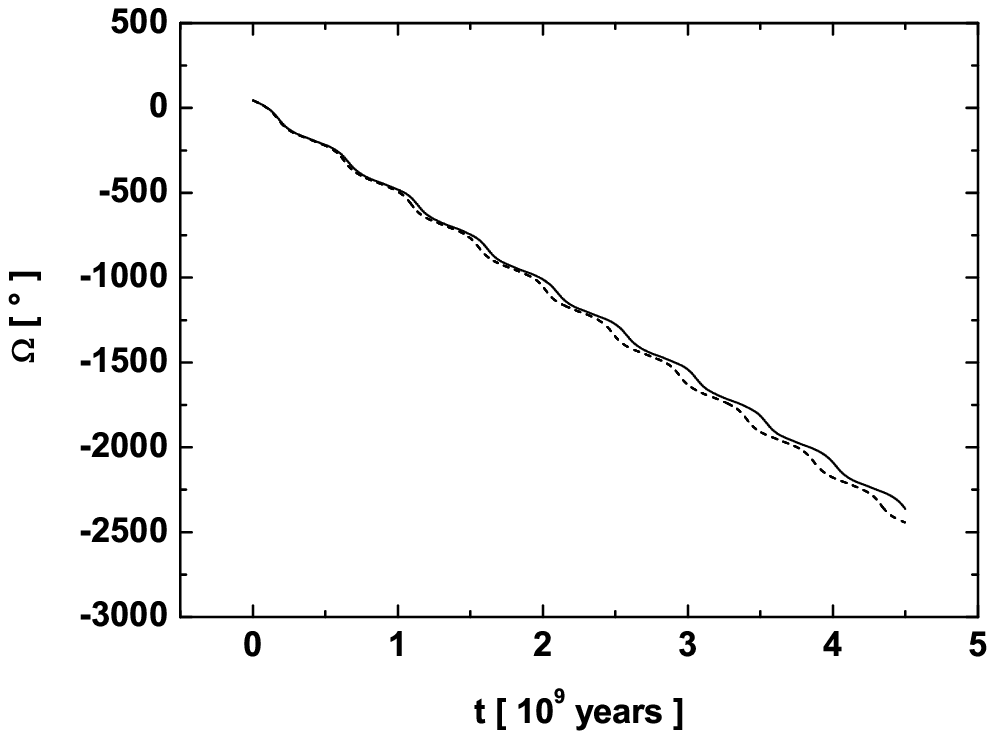}
\includegraphics[height=0.23\textheight]{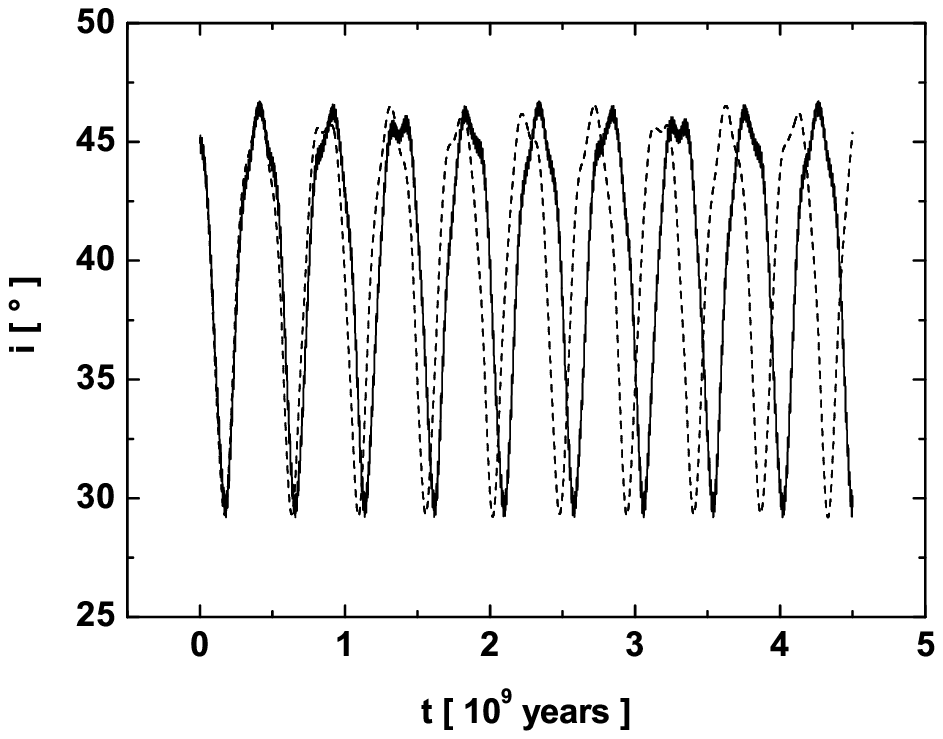}
\end{center}
\caption{Comparison of two orbital evolutions of a comet
with initial semi-major axis 50000 AU obtained by numerical
solution of Eqs. (1) (solid line) and Eqs. (13)-(17)
(dashed line). In both numerical solutions are
the new terms included.}
\label{F4}
\end{figure}

\subsection{Another method of averaging}

The procedure of averaging may not take into account time evolution
in $\omega_{0}t$ and $\omega_{z}t$ during one orbital period
if $2 \pi / \omega_{0}$ $\gg$ $T$ and $2 \pi / \omega_{z}$ $\gg$ $T$.

The secular evolution of orbital elements (and other quantitites)
can be calculated on the basis of Eqs. (4)-(5) and (8)-(10).
We define $\tilde \Omega$ as $\tilde \Omega$ $=$ $\Omega$ $+$ $\omega_{0} t$.
The final result then can be summarized in the form
\begin{eqnarray}\label{13-17}
\left \langle \frac{da}{dt} \right \rangle &=& -~ a^{2} ~
\sqrt{\frac{p}{\mu}} ~X_{a} ~Z_{0} ~\sin i ~\cos \tilde \Omega ~,
\nonumber \\
X_{a} &=&
2 ( A - B )^{2} ~ (\Gamma_{1} - \Gamma_{2} Z_{0}^{2}) ~R_{0}
~+~ 4 ~\pi ~G ~\varrho' ~,
\\
\left \langle \frac{de}{dt} \right \rangle &=&
\frac{5ae}{4} ~\sqrt{\frac{p}{\mu}} ~\bigl \{ 4A(A-B) ~\bigl [\cos i ~
\cos 2 \omega ~\sin 2 \tilde \Omega
\nonumber \\
& & + ~\sin 2 \omega ~(\cos^{2} \tilde \Omega -
\cos^{2} i ~\sin^{2} \tilde \Omega) \bigr ]
\nonumber \\
& & + ~\bigl [4 ~\pi ~G ~\varrho + 2 (A^{2}-B^{2})-(A-B)^{2} \bigr ]
~\sin^{2} i ~\sin 2 \omega \bigr \}
\nonumber \\
& & + ~5 a e ~\sqrt{\frac{p}{\mu}} ~
( A - B )^{2} ~ (\Gamma_{1} - \Gamma_{2} Z_{0}^{2}) ~R_{0} ~Z_{0} ~
\sin i ~\sin \omega
\nonumber \\
& & \times ~\bigl ( \cos i ~\cos \omega ~\sin \tilde \Omega +
\sin \omega ~\cos \tilde \Omega \bigr )
\nonumber \\
& & + ~\frac{5ae}{2} ~\sqrt{\frac{p}{\mu}} ~
4 ~\pi ~G ~\varrho' ~Z_{0} ~\sin i ~\cos \omega
\nonumber \\
& & \times ~ ( \cos ~\omega ~\cos \tilde \Omega -
\cos i ~\sin \omega ~\sin \tilde \Omega ) ~,
\\
\left \langle \frac{d \omega}{dt} \right \rangle &=&
- ~A (A - B) ~\sqrt{\frac{p}{\mu}} ~a ~\bigl [ 5 ~(\cos i ~\sin 2 \omega ~
\sin 2 \tilde \Omega
\nonumber \\
& & - ~2 \cos 2 \omega ~\cos^{2} \tilde \Omega) +
~(5 \cos 2 \omega - 3)
\nonumber \\
& & \times ~\bigl (\cos^{2} \tilde \Omega +
\frac{1}{1-e^{2}} ~\cos^{2} i ~\sin^{2} \tilde \Omega \bigr )
\nonumber \\
& & + ~\frac{\cos i}{1-e^{2}} ~\bigl ( 5e^{2} \sin 2 \omega ~
\sin \tilde \Omega ~\cos \tilde \Omega
\nonumber \\
& & - ~2 \cos i ~\sin^{2} \tilde \Omega \bigr ) \bigr ]
\nonumber \\
& & + ~\frac{a}{4(1-e^{2})} ~\sqrt{\frac{p}{\mu}} ~
\bigl [4 ~\pi ~G ~\varrho + 2 (A^{2}-B^{2})-(A-B)^{2} \bigr ]
\nonumber \\
& & \times ~\bigl [ (\sin^{2} i - e^{2})(5 \cos 2 \omega - 3) +
2 \cos^{2} i \bigr ] ~- ~\frac{3a}{2} ~\sqrt{\frac{p}{\mu}} ~(A-B)^{2}
\nonumber \\
& & - ~\frac{5 a}{1-e^{2}} ~\sqrt{\frac{p}{\mu}} ~
( A - B )^{2} ~ (\Gamma_{1} - \Gamma_{2} Z_{0}^{2}) ~R_{0} ~Z_{0} ~
\sin i ~\sin \omega
\nonumber \\
& & \times ~\bigl [ \cos i ~\sin \omega ~\sin \tilde \Omega - (1-e^{2}) ~
\cos \omega ~\cos \tilde \Omega \bigr ]
\nonumber \\
& & - ~\frac{a}{2 (1-e^{2})} ~\sqrt{\frac{p}{\mu}} ~
4 ~\pi ~G ~\varrho' ~Z_{0} ~\frac{1}{\sin i} ~
\bigl [5 ~(\sin^{2} i - e^{2}) ~\sin \omega
\nonumber \\
& & \times ~(\cos \omega ~\cos \tilde \Omega -
\cos i ~\sin \omega ~\sin \tilde \Omega)
\nonumber \\
& & + ~(1-e^{2}) ~\cos i ~\sin \tilde \Omega \bigr ] ~,
\\
\left \langle \frac{d \Omega}{dt} \right \rangle &=&
\frac{a}{4(1-e^{2})} \sqrt{\frac{p}{\mu}} ~\bigl \{ 4A(A-B) ~[ 5e^{2}
\sin 2 \omega ~\sin \tilde \Omega ~\cos \tilde \Omega
\nonumber \\
& & - ~\cos i ~(2-e^{2}(5 \cos 2 \omega - 3)) ~
\sin^{2} \tilde \Omega ]
\nonumber \\
& & - ~\bigl [4 ~\pi ~G ~\varrho + 2 (A^{2}-B^{2})-(A-B)^{2} \bigr ]
\cos i ~[2-e^{2}(5 \cos 2 \omega - 3)] \bigr \}
\nonumber \\
& & + ~\frac{a}{1-e^{2}} ~\sqrt{\frac{p}{\mu}} ~
( A - B )^{2} ~ (\Gamma_{1} - \Gamma_{2} Z_{0}^{2}) ~R_{0} ~Z_{0} ~
\sin i
\nonumber \\
& & \times ~\bigl ( 1 - e^{2} + 5e^{2} \sin^{2} \omega \bigr ) ~
\sin \tilde \Omega
\nonumber \\
& & + ~\frac{a}{2 (1-e^{2})} ~\sqrt{\frac{p}{\mu}} ~
4 ~\pi ~G ~\varrho' ~Z_{0} ~\frac{\cos i}{\sin i} ~
\bigl [ (1-e^{2}) ~\sin \tilde \Omega ~\cos i
\nonumber \\
& & - ~5e^{2} \sin \omega ~(\cos \omega ~\cos \tilde \Omega - \cos i ~
\sin \omega ~\sin \tilde \Omega) \bigr ] ~,
\\
\left \langle \frac{di}{dt} \right \rangle &=&
\frac{a}{4(1-e^{2})} ~\sqrt{\frac{p}{\mu}} ~\sin i ~
\bigl \{ 5e^{2} ~\bigl [4A(A-B) ~\sin \tilde \Omega
\nonumber \\
& & \times ~(\cos 2 \omega ~\cos \tilde \Omega- \cos i ~
\sin 2 \omega ~\sin \tilde \Omega)
\nonumber \\
& & - ~\bigl (4 ~\pi ~G ~\varrho + 2 (A^{2}-B^{2})-(A-B)^{2} \bigr ) ~
\cos i ~\sin 2 \omega \bigr ]
\nonumber \\
& & + ~2A(A-B)(2+3e^{2}) ~\sin 2 \tilde \Omega \bigr \}
\nonumber \\
& & + ~\frac{5 a e^{2}}{1-e^{2}} ~\sqrt{\frac{p}{\mu}} ~
( A - B )^{2} ~ (\Gamma_{1} - \Gamma_{2} Z_{0}^{2}) ~R_{0} ~Z_{0} ~
\sin^{2} i
\nonumber \\
& & \times ~\sin \omega ~\cos \omega ~\sin \tilde \Omega
\nonumber \\
& & - ~\frac{a}{2(1-e^{2})} ~\sqrt{\frac{p}{\mu}} ~
4 ~\pi ~G ~\varrho' ~Z_{0} ~\cos i ~\bigl [ (1-e^{2}) ~
\cos \tilde \Omega
\nonumber \\
& & + ~5e^{2} \cos \omega ~(\cos \omega ~\cos \tilde \Omega -
\cos i ~\sin \omega ~\sin \tilde \Omega) \bigr ] ~.
\end{eqnarray}

Eqs. (13)-(17) yield for the $z-$component for angular momentum per unit mass
$H_{z}$ $=$ $\sqrt{\mu ~a ~(1-e^{2})}$ $\cos i$:
\begin{eqnarray}\label{18}
\left \langle \frac{d H_{z}}{dt} \right \rangle &=&
- ~A(A-B) ~\frac{a^{2}}{2} \bigl \{ 5e^{2} \bigl [(1+ \cos^{2} i) ~
\cos 2 \omega ~\sin 2 \tilde \Omega
\nonumber \\
& & + ~2 \cos i ~\sin 2 \omega ~\cos 2 \tilde \Omega \bigr ] +
(2+3e^{2}) ~\sin^{2} i ~\sin 2 \tilde \Omega \bigr \}
\nonumber \\
& & - ~a^{2} ~
( A - B )^{2} ~ (\Gamma_{1} - \Gamma_{2} Z_{0}^{2}) ~R_{0} ~Z_{0} ~\sin i
\nonumber \\
& & \times ~\bigl [ (1-e^{2}) ~\cos i ~
\cos \tilde \Omega + 5e^{2} \sin \omega
\nonumber \\
& & \times ~( \cos i ~
\sin ~\omega ~\cos \tilde \Omega + \cos \omega ~
\sin \tilde \Omega ) \bigr ] ~.
\end{eqnarray}

Eqs. (13)-(17) show secular time derivatives of orbital elements of
a comet during one orbital period caused by the perturbation
acceleration given in Eqs. (8)-(10). The secular time derivatives,
represented by Eqs. (13)-(18), are generalization of the results obtained by
Kla\v{c}ka and Gajdo\v{s}\'{\i}k (1999) and Fouchard et al. (2005).

The secular orbital evolution holds if the method of averaging is
acceptable. The period of revolution $T$ must fulfill the condition
($\omega_{z}$ $+$ $\omega_{0}$) $T$ $\ll$ 2 $\pi$ or equivalent
condition $T$ $\ll$ ( 1 / $T_{z}$ $+$ 1 / $T_{0}$ )$^{-1}$.
This is a consequence of the terms $Z_{0} ~\sin (\omega_{0}t)$ and
$Z_{0} ~\cos (\omega_{0}t)$.

\begin{figure}[t]
\begin{center}
\includegraphics[height=0.23\textheight]{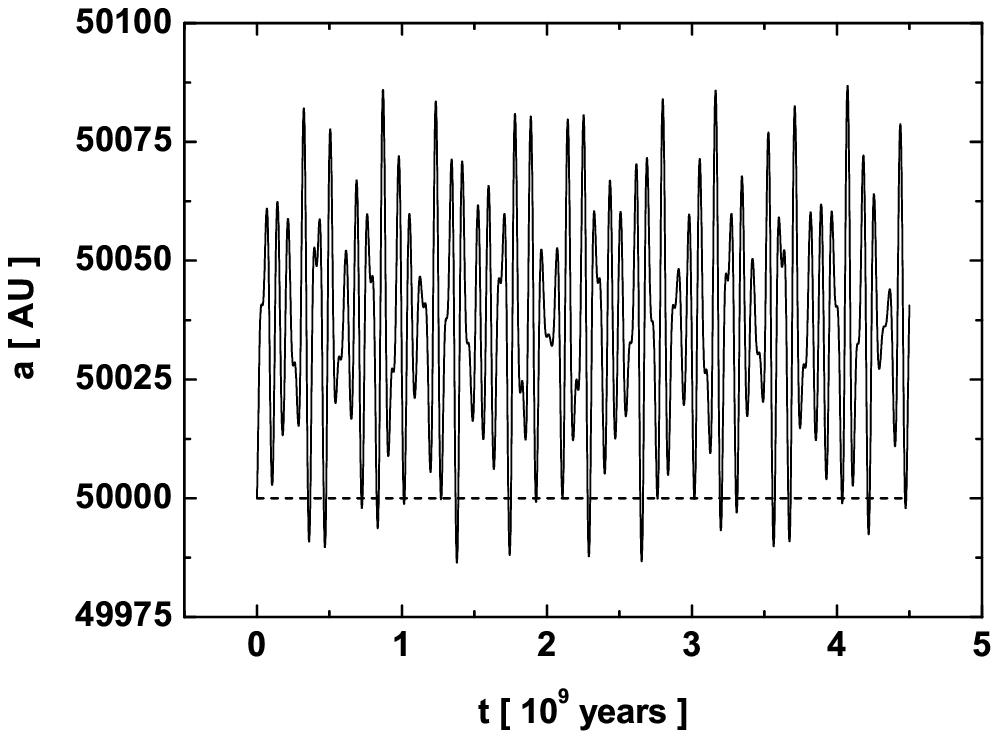}
\includegraphics[height=0.23\textheight]{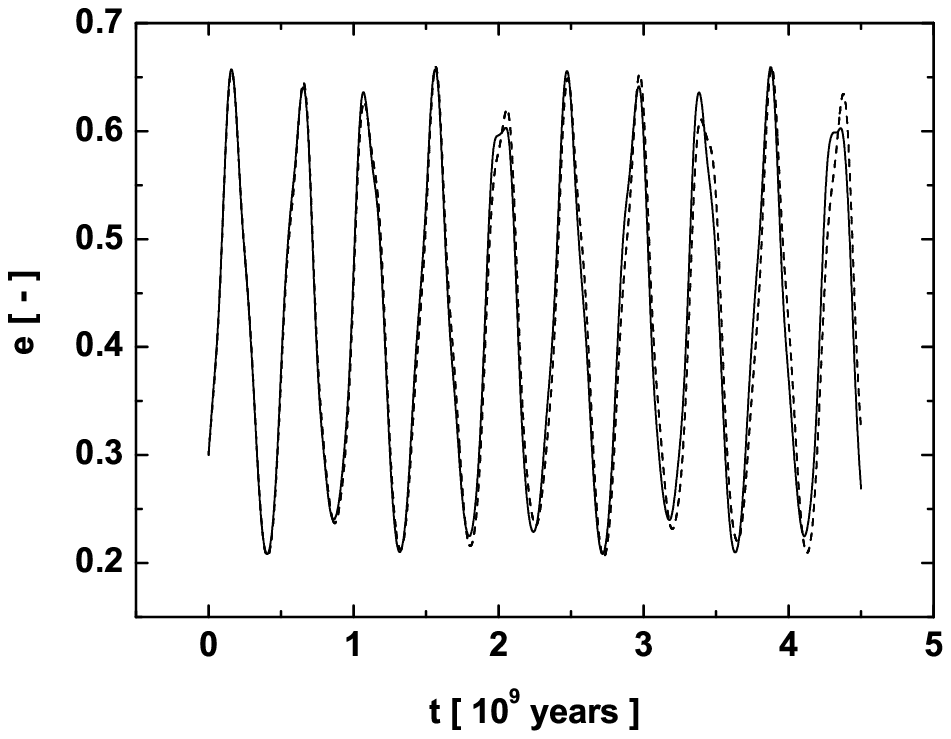}
\includegraphics[height=0.23\textheight]{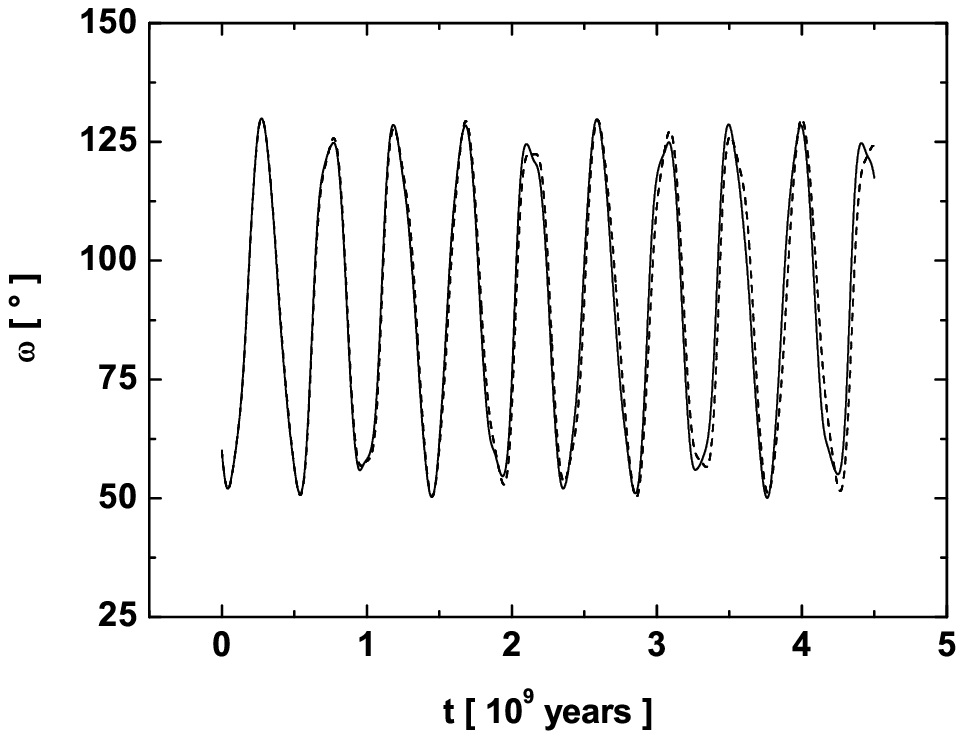}
\includegraphics[height=0.23\textheight]{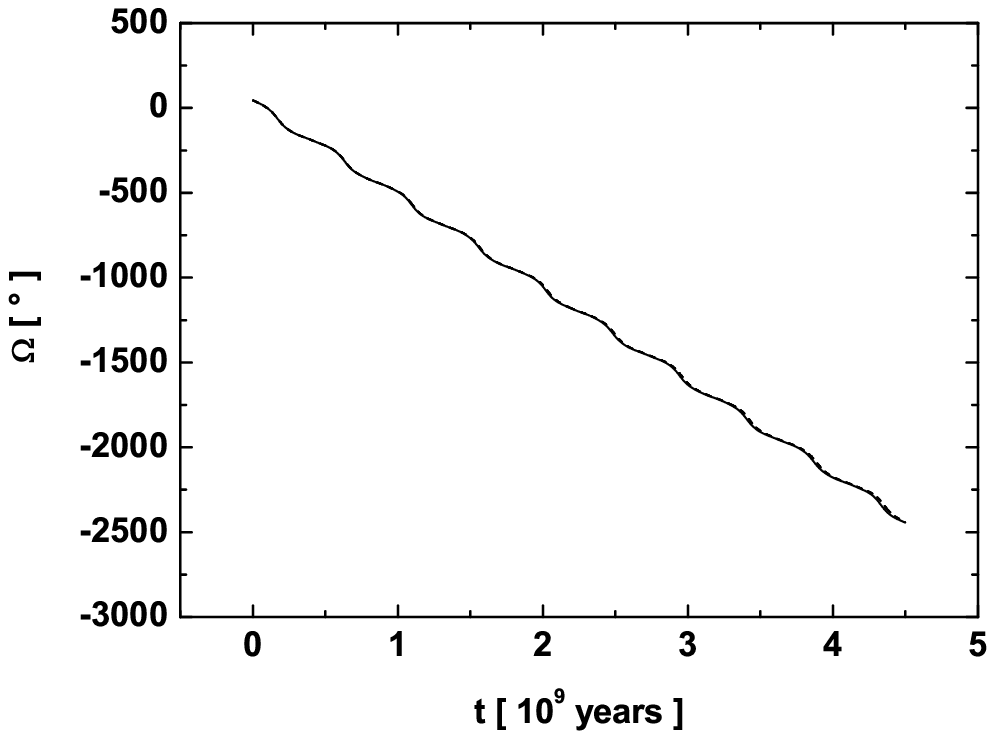}
\includegraphics[height=0.23\textheight]{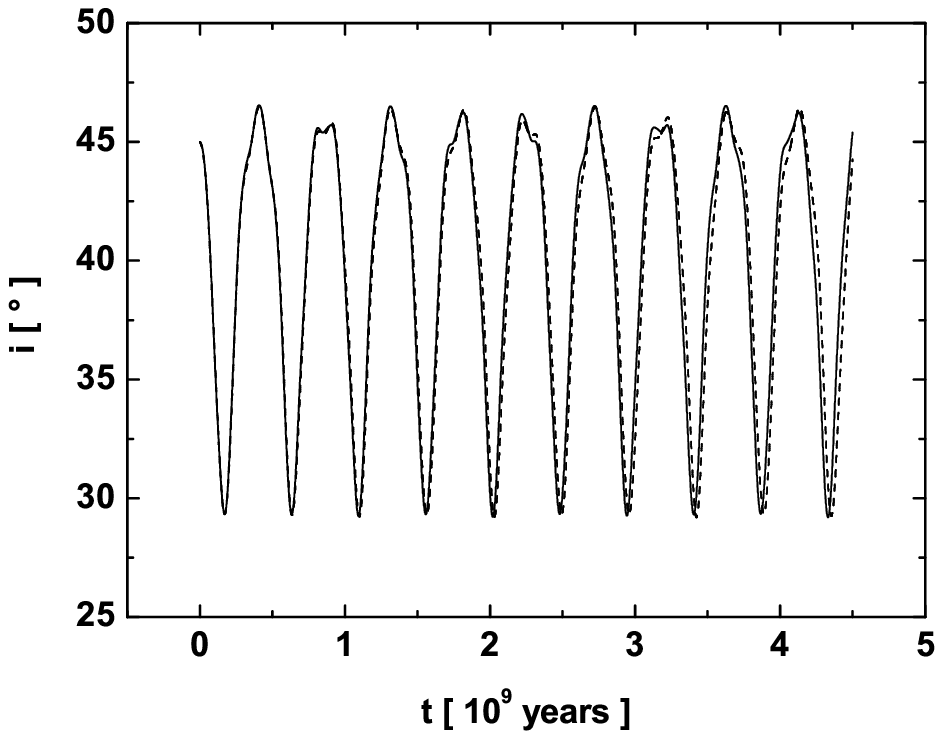}
\end{center}
\caption{Two orbital evolutions of a comet with initial semi-major axis
50000 AU obtained from numerical solution of system of differential
equations given by Eqs. (13)-(17) with the new terms (solid line)
and without new terms (dashed line).}
\label{F5}
\end{figure}

\section{Discussion}

Eqs. (13)-(17) produce identical orbital evolution
after each of the following transformations:\\
1. $\omega$ $\rightarrow$ $\omega$ + $\pi$,\\
2. $Z_{0}$ $\rightarrow$ $-$ $Z_{0}$,
$\omega$ $\rightarrow$ $\omega$ + $\pi$,
$\tilde \Omega$ $\rightarrow$ $\tilde \Omega$ + $\pi$,\\
3. $Z_{0}$ $\rightarrow$ $-$ $Z_{0}$,
$\tilde \Omega$ $\rightarrow$ $\tilde \Omega$ + $\pi$.\\
The first transformation represents a symmetry of the Oort cloud under
the action of the galactic tide. The second transformation holds
due to symmetry of the galactic potential with respect to the galactic
equatorial plane. The transformation
$\omega$ $\rightarrow$ $\omega$ + $\pi$ and
$\tilde \Omega$ $\rightarrow$ $\tilde \Omega$ + $\pi$
is equivalent to the transformation
$i$ $\rightarrow$ $-$ $i$. The third transformation can be obtained from
(simultaneous) composition of the first and the second transformation.

In order to obtain the opposite signs in time derivatives of the orbital
elements, it is sufficient to use the following transformation in
Eqs. (13)-(17):\\
$\omega$ $\rightarrow$ $\pi$ $-$ $\omega$,
$\tilde \Omega$ $\rightarrow$ $\pi$ $-$ $\tilde \Omega$.\\
This transformation represents an antisymmetry of the Oort cloud under
the action of the galactic tide.

The total secular time derivative of semi-major axis is not equal to zero.
This can have a close relation to the result represented by Eq. (31)
in Kla\v{c}ka (2009b). Non-zero value of secular time derivative
of semi-major axis is caused by new terms in our equation of motion.
The new terms contain $\Gamma_{1}$, $\Gamma_{2}$ and
$\varrho'$ quantities. Namely, the term proportional to
$\Gamma_{1} - \Gamma_{2} Z_{0}^{2}$ in $\xi$ and $\eta$ components
of the acceleration and the term proportional to $\varrho'$ in $\zeta$
component of the acceleration. Two orbital evolutions of a comet
obtained by numerical solution of Eqs. (13)-(17) are shown in
Fig. 1. Both evolutions depicted in Fig. 1 have equal initial conditions.
Initial values of the orbital elements are $a_{in}$ $=$ 10 000 AU,
$e_{in}$ $=$ 0.4, $\omega_{in}$ $=$ 0, $\Omega_{in}$ $=$ 0 and
$i_{in}$ $=$ 90$^{\circ}$. The Sun is located at distance 8 kpc from
the galactic center, $Z_{0}(0)$ $=$ 30 pc and $\dot{Z}_{0}(0)$ $=$
7.3 km s$^{-1}$ at the time $t$ $=$ 0. Evolution depicted by
a black color is calculated using secular time derivatives from
Eqs. (13)-(17) which have
$\Gamma_{1}$ $=$ 0.124 ~kpc$^{-2}$, $\Gamma_{2}$ $=$ 1.586 ~kpc$^{-4}$ and
$\varrho'$ $=$ - 0.037 M$_{\odot}$ ~pc$^{-3}$ ~kpc$^{-1}$. Evolution
depicted by a gray color is calculated from Eqs. (13)-(17) without
the new terms, i.e. by putting $\Gamma_{1}$ $=$ 0,
$\Gamma_{2}$ $=$ 0 and $\varrho'$ $=$ 0 in Eqs. (13)-(17).
Semi-major axis of the comet is not constant for the evolution
depicted by the black color in Fig. 1. Oscillation in semi-major axis
depicted by the black color is a typical behavior of semi-major axis
found from Eqs. (13)-(17) with the inclusion of the
new terms. The semi-major axis oscillates around
the value close to $a_{in}$. We did not find evolution of
semi-major axis with a tendency to a monotonic increase or decrease in time.
Only oscillations in evolution of the semi-major axis existed.
Evolutions of other orbital elements are not significantly affected
by inclusion of the new terms.

Eqs. (14) immediately show that $e$ $\equiv$ 0 if $e_{in}$ $=$ 0.
We found that if $e_{in}$ is close to zero, then the eccentricity
can increase to a value close to 1. If all other initial orbital
elements are fixed, then the time span needed for
the increase of eccentricity is usually longer for the comet with
smaller value of $e_{in}$. Such situation is depicted in Fig. 2. Three
shown evolutions differ only with the initial eccentricity of the comet.
We used the values $e_{in}$ $=$ 0.01, 0.001 and 0.0001.
The initial values of other orbital elements are $a_{in}$ $=$ 10 000 AU,
$\omega_{in}$ $=$ 45$^{\circ}$, $\Omega_{in}$ $=$ 0 and $i_{in}$
$=$ 90$^{\circ}$.

\begin{figure}[t]
\begin{center}
\includegraphics[height=0.23\textheight]{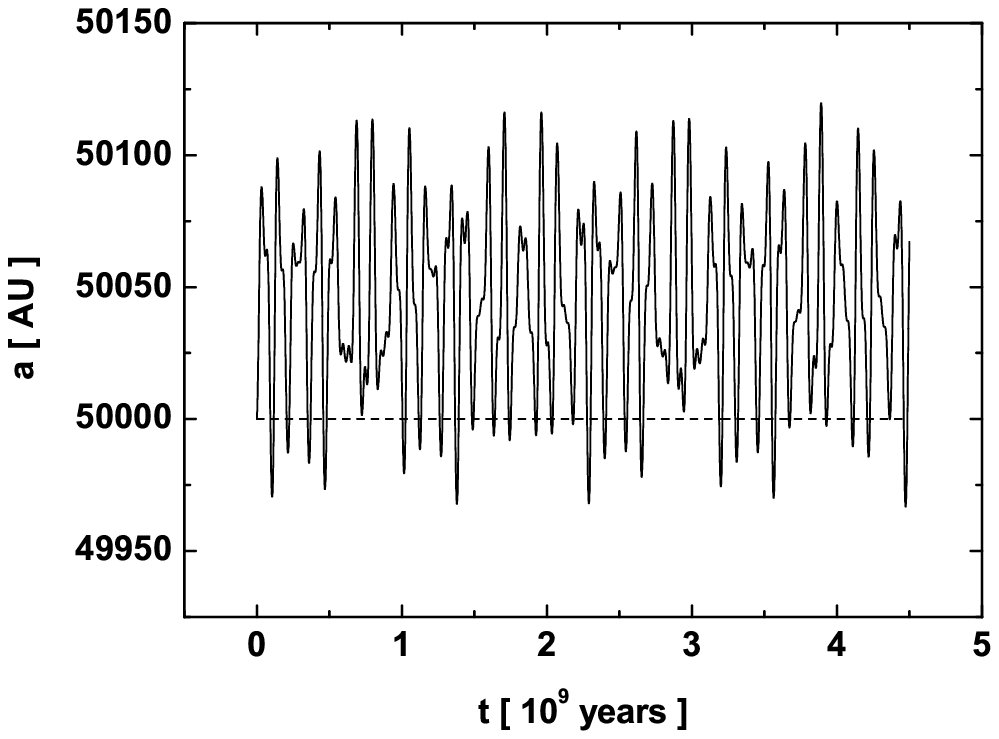}
\includegraphics[height=0.23\textheight]{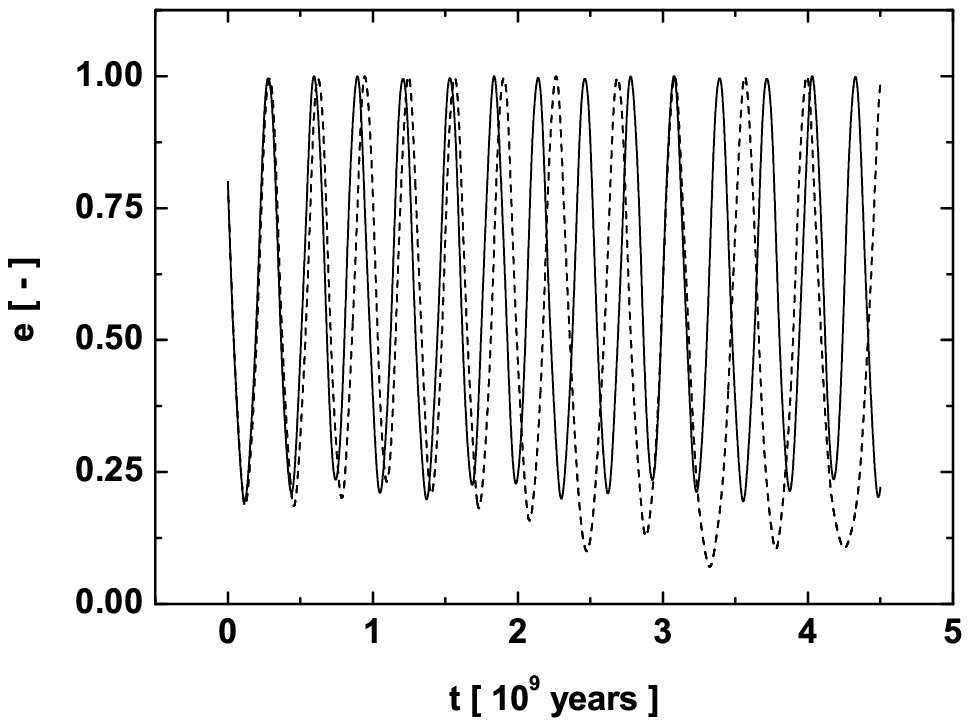}
\includegraphics[height=0.23\textheight]{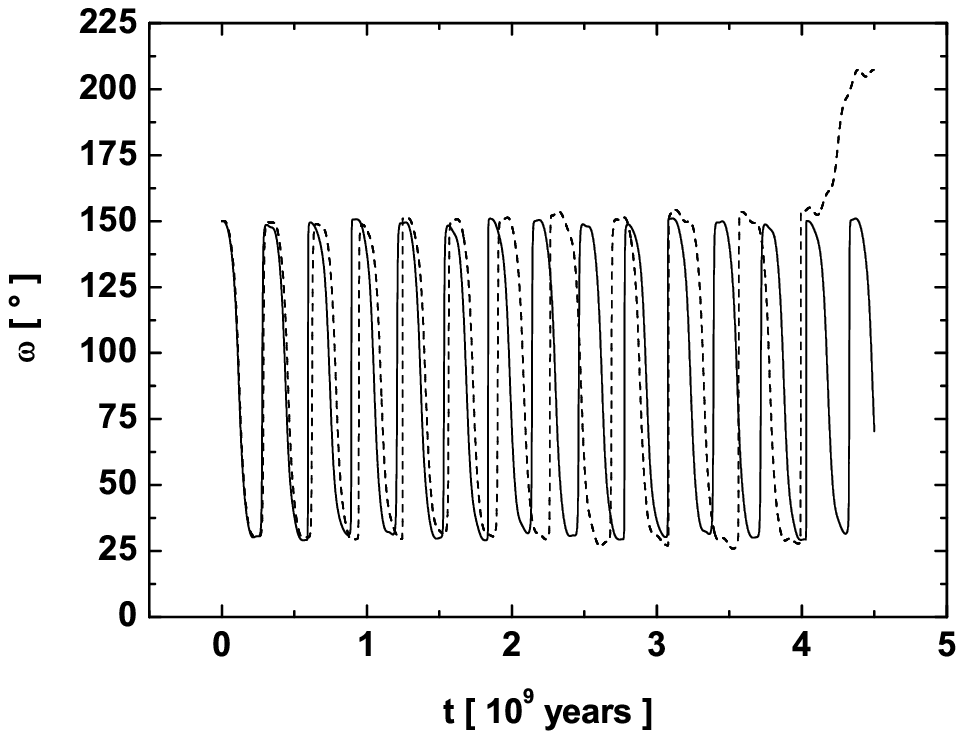}
\includegraphics[height=0.23\textheight]{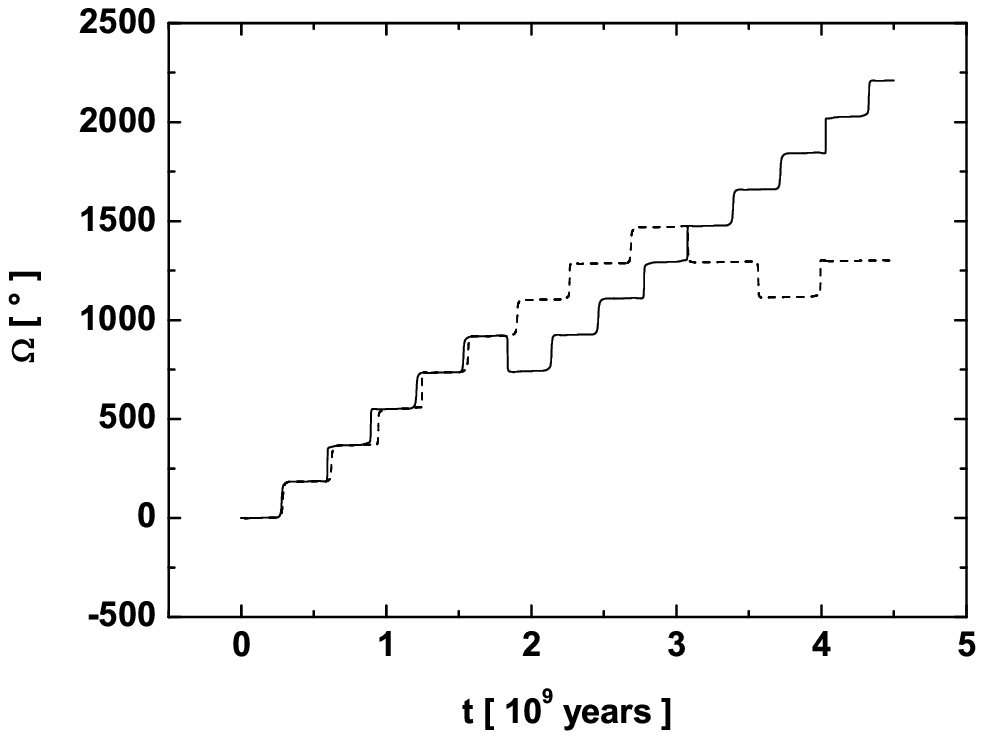}
\includegraphics[height=0.23\textheight]{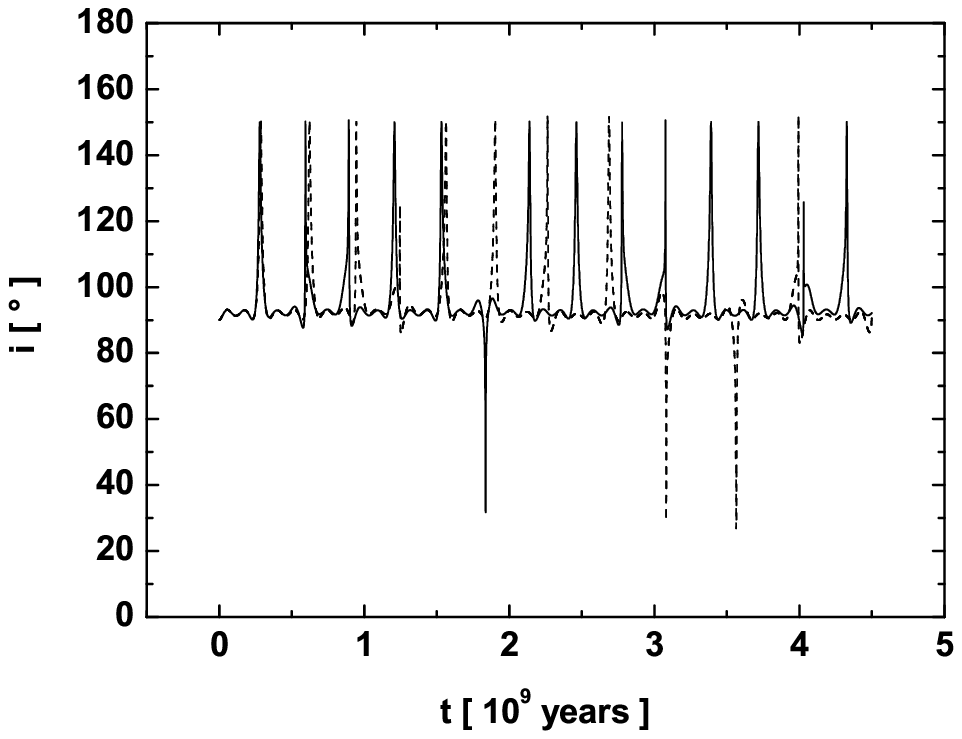}
\end{center}
\caption{Two orbital evolutions of a comet with initial semi-major axis
50000 AU obtained from numerical solution of system of differential
equations given by Eqs. (13)-(17) with the new terms (solid line)
and without new terms (dashed line).}
\label{F6}
\end{figure}

In Fig. 3 is compared solution of equation of motion Eq. (1) with
solution of system of differential equations given by Eqs. (13)-(17).
In both numerical solutions are the new term included.
Evolution depicted by a solid line is obtained from
solution of Eqs. (1) and evolution depicted by a dashed line is obtained
from solution of Eqs. (13)-(17). Initial values of
the orbital elements are $a_{in}$ $=$ 10 000 AU,
$e_{in}$ $=$ 0.3, $\omega_{in}$ $=$ 60$^{\circ}$,
$\Omega_{in}$ $=$ 45$^{\circ}$ and $i_{in}$ $=$ 45$^{\circ}$.
The Sun is located at distance 8 kpc from the galactic center,
$Z_{0}(0)$ $=$ 30 pc and $\dot{Z}_{0}(0)$ $=$ 7.3 km s$^{-1}$,
at the time $t$ $=$ 0. Initial true anomaly for the evolution
obtained from numerical solution of Eqs. (1) is $f_{in}$ $=$ 0.
Orbital evolutions obtained from numerical solutions are in good
accordance.

We used semi-major axis $a_{in}$ $=$ 50000 AU for a comparison of numerical
solutions of Eq. (1) and Eqs. (13)-(17) with the new terms at greater
semi-major axes. Initial values of other cometary orbital elements
and solar initial position and velocity are the same as for the evolutions
depicted in Fig. 3. Results are depicted in Fig. 4. Difference between
two evolutions is caused by the large value of the initial semi-major
axis of the comet. The initial semi-major axis is so large that
the orbital period of the comet ($T$ $\approx$ 1.1 $\times$ 10$^{7}$ years)
is comparable with the period of oscillations of the Sun around
the galactic equator (2$\pi/\omega_{z}$ $\approx$ 7.3 $\times$ 10$^{7}$
years). Secular orbital evolution given by Eqs. (13)-(17) cannot be used in
this case.

We compared also influence of the new terms on secular evolution
of orbital elements at larger semi-major axis for the initial conditions
corresponding to those used in Fig. 4. The resulting evolutions
are presented in Fig. 5. Evolution depicted by a solid line is for
Eqs. (13)-(17) with the new terms and evolution depicted by a dashed
line is for Eqs. (13)-(17) without the new terms. Evolution depicted
by the solid line in Fig. 5 is identical to the evolution depicted
by the dashed line in Fig. 4. Comparison of Figs. 5 and 4 shows that
the influence of the new terms in Eqs. (13)-(17) on orbital evolution
is less significant than motion of the Sun which was neglected
in derivation of Eqs. (13)-(17).

Fig. 6 depicts numerical integration of Eqs. (13)-(17) for the case
when inclusion of the new terms plays a relevant role.
The new terms significantly changed the cometary orbital evolution
in comparison with the orbital evolution without the new terms.
Both evolutions depicted in Fig. 6 had equal initial conditions.
Initial values of the comet's orbital elements were $a_{in}$ $=$ 50 000 AU,
$e_{in}$ $=$ 0.8, $\omega_{in}$ $=$ 150$^{\circ}$, $\Omega_{in}$ $=$ 0 and
$i_{in}$ $=$ 90$^{\circ}$. The Sun was located at distance 8 kpc from
the galactic center, $Z_{0}(0)$ $=$ 30 pc and $\dot{Z}_{0}(0)$ $=$
7.3 km s$^{-1}$, at the time $t$ $=$ 0. Influence of the new terms
can be even more significant for larger semi-major axis of the comet.

\section{Conclusion}

The paper treats the effect of the galactic tide on motion of a comet
with respect to the Sun. It turns out that the important effect
from the galactic tide is the action of the normal $z-$component.
The $x-$ and $y-$ components of the acceleration comes not only
from the $x-$ and $y-$ positional components of the comet, but also from
the $z-$component of the position. This is generated by the galactic disk.
The effect of the three positional components in the
$x-$ and $y-$ acceleration components are of comparable importance.

The inclusion of the new terms into the equation
of motion of the comet leads to orbital evolution which may
significantly differ from the conventional result. This is true mainly
for the comets with large semi-major axes. The conventional result
(see, e. g. Fouchard et al. 2008) is obtained from Eqs. (13)-(17) putting
$\Gamma_{1}$ $=$ 0, $\Gamma_{2}$ $=$ 0 and $\varrho'$ $=$ 0.

The solution of Eqs. (13)-(17) is in a good agreement with
the solution of the equation of motion represented by Eqs. (1), if
$T$ $\ll$ ( 1 / $T_{z}$ $+$ 1 / $T_{0}$ )$^{-1}$.

We found that a comet with the argument of perihelion $\omega$
has the same orbital evolution as a comet with the argument of perihelion
$\omega$ $+$ $\pi$, if values of other orbital elements are equal.
Similarly, a comet with the argument of perihelion $\omega$
and the longitude of the ascending node $\Omega$ has exactly opposite time
derivatives of the orbital elements as a comet with the argument
of perihelion $\pi$ $-$ $\omega$ and the longitude of ascending
node $\pi$ $-$ $\Omega$, if values of other orbital elements are equal.

\section*{Acknowledgement}
This work was supported by the Scientific Grant Agency VEGA, Slovak Republic,
grant No. 2/0016/09.


\begin{thebibliography}{9}

\bibitem{}Fouchard, M., Froeschl\'{e}, C., Breiter, S., Ratajczak, R.,
Valsecchi, G.B., Rickman, H.: Methods for the study of the dynamics of
the Oort cloud comets II: Modelling the galactic tide. Lect. Notes Phys.
729, 273-296 (2008)

\bibitem{}Fouchard, M., Froeschl\'{e}, C., Matese, J.J., Valsecchi, G.:
Comparison between different models of galactic tidal effects on
cometary orbits. Celest. Mech. and Dynam. Astron.
93, 229-262 (2005)

\bibitem{}Kla\v{c}ka, J.: Electromagnetic radiation and motion of
a particle. Celest. Mech. and Dynam. Astron. 89, 1-61 (2004)

\bibitem{}Kla\v{c}ka, J.: Galactic tide. arXiv:astro-ph/0912.3112 (2009a)

\bibitem{}Kla\v{c}ka, J.: Galactic tide in a noninertial frame of reference.
arXiv:astro-ph/0912.3114 (2009b)

\bibitem{}Kla\v{c}ka, J., Gajdo\v{s}\'{\i}k, M.:
Orbital motion in outer Solar System.
In: Pretka-Ziomek, H., Wnuk, E., Seidelmann, P.K., Richardson, D. (eds.)
Dynamics of Natural and Artificial Celestial Bodies, pp. 347-349.
Kluwer Academic Press, Dordrecht, arXiv:astro-ph/9910041 (2001)

\bibitem{}K\'{o}mar, L., Kla\v{c}ka, J., P\'{a}stor, P.: Galactic tide and
orbital evolution of comets. arXiv:astro-ph/0912.3447 (2009)

\end{thebibliography}
\end{document}